\documentclass[journal]{IEEEtran}
\IEEEoverridecommandlockouts
\newcommand{\subparagraph}{}
\usepackage[american,fulldiode]{circuitikz}
\usepackage{epstopdf}
\usepackage{caption}
\usepackage[linesnumbered,ruled,vlined]{algorithm2e}

\usepackage[cmex10]{amsmath}
\usepackage{array}
\usepackage{mdwmath}
\usepackage{mathrsfs}
\usepackage{mdwtab}
\usepackage{caption}
\usepackage[caption=false,font=footnotesize]{subfig}
\usepackage{cite}
\usepackage{amssymb}
\renewcommand\Re{\operatorname{Re\mathfrak{}}}

\usepackage{blindtext, subfig}
\usepackage{multicol}
\usepackage{amssymb}

\mathcode"65"65

\usepackage{amssymb}
\usepackage{amsmath,scalefnt}
\usepackage{mathtools}

\def\bad{\spaceskip=0.33emplus0.6emminus0.15em\immediate\write5{\string\bad}}

\def\({\left(}
\def\){\right)}
\def\Re{\operatorname{Re}}

\def\mcap{\operatorname{cap}}

\def\RR{\mathbb R}
\def\CC{\mathbb C}

\def\HH{\mathscr H}

\let\geq\geqslant
\let\ge\geqslant
\let\myo\overline
\def\bad{\spaceskip=0.33emplus0.6emminus0.15em\immediate\write5{\string\bad}}

\def\bad{\spaceskip=0.33emplus0.6emminus0.15em\immediate\write5{\string\bad}}

\def\({\left(}
\def\){\right)}
\def\Re{\operatorname{Re}}

\def\mcap{\operatorname{cap}}

\def\RR{\mathbb R}
\def\CC{\mathbb C}

\def\HH{\mathscr H}

\let\geq\geqslant
\let\ge\geqslant
\let\myo\overline
\def\bad{\spaceskip=0.33emplus0.6emminus0.15em\immediate\write5{\string\bad}}

\begin{document}

\title{Embedding AC Power Flow in the Complex Plane Part I: Modelling and Mathematical Foundation}



\author{Sina S. Baghsorkhi, \textit{Member, IEEE}, and Sergey P. Suetin\vspace{-2mm}}

\maketitle
\pagenumbering{gobble}

\begin{abstract}
Part I of this paper embeds the AC power flow problem with voltage control and exponential load model in the complex plane. Modeling the action of network controllers that regulate the magnitude of voltage phasors is a challenging task in the complex plane as it has to preserve the framework of holomorphicity for obtention of these complex variables with fixed magnitude. The paper presents two distinct approaches to modelling the voltage control of generator nodes. Exponential (or voltage-dependent) load models are crucial for accurate power flow studies under stressed conditions. This new framework for power flow studies exploits the theory of analytic continuation, especially the monodromy theorem for resolving issues that have plagued conventional numerical methods for decades. Here the focus is on the indispensable role of Pad\'{e} approximants for analytic continuation of complex functions, expressed as power series, beyond the boundary of convergence of the series. The zero-pole distribution of these rational approximants serves as a proximity index to voltage collapse. Finally the mathematical underpinnings of this framework, namely the Stahl's theory and the rate of convergence of Pad\'{e} approximants are explained.
\end{abstract}

\begin{IEEEkeywords}
AC power flow, voltage control, exponential load model, power flow feasibility, saddle-node bifurcation, algebraic curves, analytic continuation, monodromy, continued fractions, Pad\'{e} approximants, reduced Gr\"{o}bner basis, Stahl's compact set. \vspace{0mm}
\end{IEEEkeywords}

\section{Introduction}

Power flow, the most fundamental concept in power system engineering, is at the heart of studies ranging from daily operation to long-term planning of electricity networks. The AC power flow problem is a system of nonlinear algebraic equations that mathematically models the steady-state relations between the phasor representation of parameters and unknown states in an AC circuit. The parameters typically consist of the power generated and consumed by source and sink nodes and the electrical properties, i.e. the impedance, of lines that connect these nodes. The unknown states are primarily voltage phasors but could also include continuous or discrete variables associated with network controllers, e.g. FACTS devices and tap-changing or phase-shifting transformers. The accurate and reliable determination of these states is imperative for control and thus for efficient and stable operation of the network. In certain studies it is equally vital to determine for which parameter values the power flow problem becomes infeasible as this condition is intimately linked to saddle-node bifurcation and the voltage collapse phenomenon where the system loses structural stability~\cite{Venikov, Sauer, VC}. The concept of structural stability is only applicable to dynamical systems~\cite{Andronov,Arnold}. However as demonstrated previously in the context of the differential-algebraic equations that model a power system, the analysis of the static model, i.e. the algebraic equations, is sufficient to determine where exactly in the parameter space the system loses structural stability. The dynamical model of load and generators are only needed to capture oscillatory instability phenomena, such as Hopf bifurcation, that can arise from the interaction of generators and their controllers with the network~\cite{Dobson}. Thus the distance in the parameter space to power flow infeasibility {\em might be}\footnote{ As demonstrated in Part II of this paper, the power flow feasibility and saddle-node bifurcation boundaries are not necessarily equivalent. In that case, the power flow would be still feasible, albeit with a non-physical solution, beyond the closest saddle-node bifurcation.} regarded as the margin of voltage stability~\cite{VC}. Voltage collapse and bifurcation is certainly one of the most theoretical areas in electrical engineering. As conventional power systems undergo a fundamental transformation by large-scale highly-variable wind and solar generation distributed across the network, this field may experience a resurgence~\cite{SB}. Given the inherent limitations of traditional methods, deeper understanding of these complicated phenomena requires new theoretical approaches rooted in complex analysis and algebraic geometry.

The basis of power flow is Kirchhoff's current law which states that for every node $i$ in $\mathcal{N}$, the set of all nodes, $I_i$, the net current flowing out of that node, is related to its voltage $V_i$ and those of its adjacent nodes $V_k$ in the following way:\vspace{-2mm}

\begin{align}\label{}
\displaystyle I_i=\sum_{k\in \mathcal{N}(i)}I_{ik}= \sum_{k\in \mathcal{N}(i)} \frac{V_i-V_k}{Z_{ik}} =\sum_{k\in \mathcal{N}[i]} V_kY_{ik}
\end{align}

$\mathcal{N}(i)$ and $\mathcal{N}[i]$ are the open and closed neighborhoods of node $i$. $I_{ik}$ is the current flow through the line connecting node $i$ and $k$ and $Z_{ik}=R_{ik}+jX_{ik}$ is the impedance of that line which is used to construct the diagonal and off-diagonal elements of the admittance matrix as,\vspace{-3mm}

\begin{align}\label{}
\displaystyle Y_{ii}=\sum_{k\in \mathcal{N}(i)} \frac{1}{Z_{ik}}, \hspace{2mm} Y_{ik}=-\frac{1}{Z_{ik}}
\end{align}

Since complex power is $S_i=P_i+jQ_i=V_iI_i^*$, the power flow problem in its complex form can be expressed as,
\begin{align}\label{}
&\displaystyle S^*_i=\sum_{k\in \mathcal{N}[i]} V^*_iV_k Y_{ik} & \forall i\in  \mathcal{N}-\{r\} \label{complex}
\end{align}

Here $r$ is the voltage reference node with $|V_r|=\text{constant}$ and $\arg(V_r)=0$. It also serves as the slack node meaning that $S_r$ is a free parameter that accounts for the mismatch of complex power and its losses throughout the network.

The numerical methods, developed historically to solve this problem, take the polynomial system of (\ref{complex}) out of its complex form by reformulating it in rectangular or polar forms. These techniques, all based on Newton's method or its variants, iteratively linearize the equations and approximate the solution, starting from an initial guess.

There are two inherent shortcomings in such methods that can arise near the feasibility boundary of~\eqref{complex} characterized by the saddle-node bifurcation manifold in its parameter spaces. Physically, proximity to the feasibility boundary corresponds to a network operating close to its loadability limit such as periods of peak electricity demand. The first issue is the increased likelihood of non-convergence even though the operating point is still feasible. The second issue near the feasibility boundary where different algebraic branches coalesce is convergence to solutions that lie on other algebraic branches. Although dependent on the dynamical model of the physical system, these solutions in power systems typically signify unstable~\cite{VC} or low voltage~\cite{Sauer2} operating points. Most of these operating points cannot be physically realized and are thus false solutions. The region of initial guesses in Newton's method that converge to a particular solution has a fractal boundary~\cite{Thorp}. The multiple fractal domains of convergence are pressed together near the bifurcation manifold which explains erratic behavior of such methods in finding the desirable, i.e. stable, solution even with seemingly reasonable initial guesses.

Recently a semidefinite relaxation of rectangular power flow has been reformulated as a special case of optimal power flow where the objective function of the semidefinite programming (SDP) is minimizing active power loss~\cite{DM}. This addresses the convergence failure of iterative methods but has its own serious drawbacks. First, the relaxation may not be tight and yield a high rank matrix where it is impossible to recover any solution to the original power flow, let alone the desirable one. Second, if the solution of the relaxed problem is high rank, nothing can be concluded on the feasibility of the power flow in the same vein as non-convergence of iterative methods cannot rule out the existence of solutions. Third, the suggested heuristic, i.e. active power loss minimization, does not always find the stable solution branch. The first two problems can be remedied, at least in theory, by obtaining higher-order and thus tighter relaxations of power flow equations. The computation cost however,  explodes with the order of relaxation and the number of variables. Reference~\cite{Lasserre} discusses the theoretical underpinning of this approach in the context of the generalized moment problem and highlights its connection to {\em real algebraic geometry} which we see as an obstacle to distinguishing the desirable solution branch for algebraic systems.

Among the above issues, the challenge of finding the solution on the desirable branch, more than anything else, underlines the significance of embedding the power flow problem in the complex plane where the extraordinary potentials of analytic continuation theory for multi-valued complex functions can be tapped. This is pioneered by the idea of holomorphic embedding load flow (HELM) which builds on the fact that under no load/no generation ($S_i=0 \hspace{2mm} \forall i\in  \mathcal{N}$), the network has a trivial non-zero solution for voltage phasors. This corresponds to all currents $I_{ik}$ being zero and reference voltage $V_r$ propagated across the network and this trivial solution characterizes the stable branch\footnote{These concepts are explained more concretely in Part II of this paper.}. Analytic continuation of this solution (or more accurately speaking, the germs developed at $z=0$) is guaranteed by monodromy theorem to yield the desirable solution all the way to the saddle-node bifurcation in the parameter space of~\eqref{complex} where there is a non-trivial monodromy (Appendix) and the physical solution ceases to exist.

Although the idea of HELM has aroused significant interest in the power system community, it still needs much further development before it can prove its superiority over conventional methods. Here we demonstrate how the magnitude of power flow complex variables, considered as functions of a single complex variable $z$, can be held fixed while preserving the framework of holomorphicity. This is an important step in the embedding of power flow as it models the action of network controllers in the complex plane. Under stressed conditions, i.e. near the feasibility boundary, voltage magnitudes tend to deviate far below their nominal values. Hence the modelling of voltage-dependent or more generally exponential load is another crucial aspect that is developed here. We also show the indispensable role of Pad\'{e} approximants in the cases of voltage control and exponential load models. Throughout the paper we refer to this method as PA to highlight the central role of rational approximation of functions of a complex variable for recovering the power flow solution. With this abbreviation we also wish to emphasize the critical direction of research and potential challenges for further development of this method.

The Part I of this paper is organized as follows. In Section II we review the main ideas of HELM as presented in the original paper~\cite{Trias}, i.e. for the PQ buses, introduce the concept of rational approximation of analytic functions in relation to power series and continued fractions and explore the zero-pole structure of Pad\'{e} approximants for a 3-bus example. In Section III we introduce the mathematical {\em static} model of the most prevalent controller in the network, the automatic voltage regulator (AVR) of the generator. We demonstrate through modification of the previous 3-bus network, this time with a generator (PV bus), how the approximation of functions of a single complex variable is essential for analytic continuation of the voltage phasors with fixed magnitude. We start from the analysis of the parameterized polynomial system of equations and obtain their corresponding algebraic curves via reduced Gr\"{o}bner basis method. We obtain the critical points of these curves, interpret the zero-pole structure of Pad\'{e} approximants and its transformation as the solution reaches the feasibility (or bifurcation) boundary and explain the significance of the zero-pole distribution of the Pad\'{e} approximants in terms of voltage stability margin at a given operating point. In Section IV we introduce an alternative approach for modelling the voltage magnitude constraints. In Section V we address the shortcomings of a previous formulation in the literature that attempts to incorporate the PV buses into the general framework of HELM. In section VI we introduce the exponential load model and its special case the ZIP load. In the Appendix we explain the mathematical underpinning of this paper including the concept of germ and the monodromy of multi-valued algebraic functions in relation to the Stahl's theory. We also explain the rate of convergence of Pad\'{e} approximants in the context of the Stahl's maximal domain and its Green's function. As illustrated in Section III, the method of embedding the equations and the resulting structure of the analytic arcs (branch cuts) has implications for the rate of convergence of Pad\'{e} approximants and may hinder the effective analytic continuation of the developed germs. The numerical values of power flow variables and parameters presented in this paper are all in per unit.

\section{Embedding the System of Equations in the Complex Plane}

Consider the following parametrization of\vphantom{~\eqref{complex}} complex power flow equations in terms of $z\in \mathbb{C}$ with $V^*_{i}$ replaced with independent variables $W_i$,\vspace{-1mm}
\begin{subequations}\label{polynomial}
\begin{align}
\label{poly1}
&\displaystyle zS^*_i=\sum_{k\in \mathcal{N}[i]}W_i V_k Y_{ik} &\forall i\in  \mathcal{N}-\{r\}\\
\label{poly2}
&\displaystyle zS_i=\sum_{k\in \mathcal{N}[i]} V_i W_k Y^*_{ik} & \forall i\in  \mathcal{N}-\{r\}
\end{align}
\end{subequations}

\vspace{-1mm}From a geometric point of view, the $2n$ equations of~\eqref{polynomial} define {\em generically} an affine algebraic curve in $(z,V_1,V_2,...,W_n)$. It follows from the Kirchhoff's current law and the existence of the voltage reference node (with $V_r$ appearing in~\eqref{polynomial} as a parameter) that the polynomials on the right side of~\eqref{poly1}-\eqref{poly2} (i.e. $\sum_{k\in \mathcal{N}[i]}W_i V_k Y_{ik}$ and $\sum_{k\in \mathcal{N}[i]} V_i W_k Y^*_{ik}$ $\forall i\in  \mathcal{N}-\{r\}$) are algebraically independent. To establish this algebraic independence in relation to the reference node requires rigorous analysis, a task which lies outside the scope of this paper. Taking the algebraic independence of these polynomials for granted, degenerate cases where the equations of~\eqref{polynomial} define not an algebraic curve but a higher-dimension algebraic variety can only arise when the power flow problem is ill-defined as in the case of networks with disconnected graphs. This is in line with the physical intuition that in the absence of a reference voltage, voltages are floating and a given $V_i$ can assume any value in $\mathbb{C}$. The equations of~\eqref{polynomial} generate an ideal and give a starting basis for finding the corresponding reduced Gr\"{o}bner basis~\cite{Cox}. For any lexicographic order, such as $...>V_i>z$, this gives a last basis element which is a bivariate polynomial $f_i(V_i,z)$ for well-defined problems. $f_i(V_i,z)=0$ can be solved for an algebraic (multi-valued) function $V_i = V_i(z)$ which has holomorphic branches where $\partial{f_i(V_i,z)}/\partial{V_i}\neq 0$. By permuting the order, we arrive at $2n$ algebraic functions $V_i = V_i(z)$, $W_i = W_i(z)$, $i = 1, . . . , n$, giving an algebraic parametrization of the curve defined by~\eqref{polynomial}. A $z$-critical point of this curve will be where any of the components $V_i(z)$ or $W_i(z)$ has a branch point, i.e., $f_i(V_i,z)=0$ and $\partial{f_i(V_i,z)}/\partial{V_i}=0$ (similarly for the $W_i$)~\cite{Geom}. The branch point closest to the point $z_0$ at which the Taylor series expansion of any single-valued branch is developed, determines the radius of convergence of the series. Branch points play a critical role in the analytic continuation of these functions and the PA method (Appendix). This analysis also extends to the case of voltage control where we introduce new functions $\overline{V}_i(z)$, $Q_i(z)$ and $S_i(z)$.

Since $V_i(z)$ is analytic in $z$,  $(V_i(z^*))^*$ is also analytic in $z$ and identical to the conjugate of $V_i(z)$ on the real axis. Hence the solution process involves analytic continuation of the functions of a single complex variable from $z=0$ to $z=1$ in the following system,\vspace{-1mm}

\begin{align}
\label{p}
&\displaystyle \frac{zS_i^*}{(V_i(z^*))^*}=\hspace{-2mm}\sum_{k\in \mathcal{N}[i]} V_k(z) Y_{ik} \hspace{20mm} \forall i\in\mathcal{N}-\{r\}
\end{align}

By defining $V_i(z)=\sum_{n=0}^{\infty} c_n^{[i]}z^n$, $1/V_i(z)=\sum_{n=0}^{\infty} d_n^{[i]}z^n$ and $1/ (V_i(z^*))^*=\sum_{n=0}^{\infty} {d^*_n}^{[i]}z^n$, and requiring that $(V_i(z^*))^*\neq0$ this system is adequately described by the following set of power series relations,\vspace{-1mm}

\begin{subequations}\label{HELM1}
\begin{align}
\label{main}
&zS^*_i\sum_{n=0}^{\infty} {d^*_n}^{[i]}z^n=\hspace{-3mm}\sum_{k\in \mathcal{N}[i]}\hspace{-2mm}(Y_{ik}\sum_{n=0}^{\infty} {c_n}^{[k]}z^n) \hspace{2.5mm} \forall i\in  \mathcal{N}-\{r\}\\
\label{conv}
& \displaystyle (\sum_{n=0}^{\infty} c_n^{[i]}z^n)(\sum_{n=0}^{\infty} d_n^{[i]}z^n)=1 \hspace{18mm} \forall i\in  \mathcal{N}-\{r\}
\end{align}
\end{subequations}

The procedure to obtain the coefficients of~\eqref{HELM1} starts by setting $z=0$ in~\eqref{main}. This gives the linear system of $ \sum_{k\in \mathcal{N}[i]}Y_{ik}{c_0}^{[k]}=0$ which always yields the trivial solution (i.e. $c_0^{[i]}=V_{r} \hspace{2mm} \forall i\in  \mathcal{N}-\{r\}$). Next $d_0^{[i]}=1/c_0^{[i]}$ by setting $z=0$ in~\eqref{conv}. The higher order coefficients are progressively obtained by solving the linear system of~\eqref{ln} $(\forall i\in  \mathcal{N}-\{r\})$ which itself is obtained by differentiating~\eqref{main} with respect to $z$ and evaluating at $z=0$ and the convolution formula of~\eqref{conv2}.\vspace{-1mm}
\begin{subequations}\label{HELM2}
\begin{align}
\label{ln}
&S^*_i{d^*_{n-1}}^{[i]}=\sum_{k\in \mathcal{N}[i]}Y_{ik}{c_n}^{[k]}  \hspace{15mm} \forall i\in  \mathcal{N}-\{r\}\\
\label{conv2}
&\displaystyle d_n^{[i]}=\frac{-\sum_{m=0}^{n-1}c_{n-m}^{[i]}d_m^{[i]}}{c_0^{[i]}} \hspace{18.5mm} \forall i\in  \mathcal{N}-\{r\}
\end{align}
\end{subequations}

The radius of convergence is $R=\lim_{n\rightarrow\infty}|c_n|/|c_{n+1}|$, if the limit exists. This marks the distance from the origin to the closest branch point. Notice that when an analytic function does not have a closed-form expression as in this case, its representation as a power series expansion can be approximated by a partial sum of a finite order. Since this approximation for $V_i(z)$ does not converge for $|z|\ge R$ the analytic continuation of these complex functions toward $z=1$ requires an alternative representation of these analytic functions. One such representation, with superior convergence properties, is a continued $C$-fraction which is approximated by truncation. The relation between these two representations is crucial for understanding Pad\'{e} approximants and is described below~\cite{BaGr96},

For a given power series $V(z)=c_0+c_1z+c_2z^2+...$, assume the existence of the reciprocal relation between the original series, as modified below, and a new series indexed by superscript $^{(1)}$,\vspace{-1mm}
\begin{align}\label{}
\displaystyle 1+\frac{c_2z}{c_1}+\frac{c_3z^2}{c_1}+...= (1+c_1^{(1)}z+c_2^{(1)}z^2+...)^{-1}
\end{align}

Now the original power series can be expressed as,
\vspace{-1mm}
\begin{align}\label{cf1}
\displaystyle c_0+c_1z+c_2z^2+...=c_0+ \displaystyle \frac{c_1^{(0)}z}{1+c_1^{(1)}z+c_2^{(1)}z^2+...}
\end{align}

Next assume the existence of another reciprocal relation between the modified series from the denominator of the fraction in~\eqref{cf1} and a new series indexed by superscript $^{(2)}$,
\vspace{-1mm}
\begin{align}\label{}
\displaystyle1+\frac{c_2^{(1)}z}{c_1^{(1)}}+\frac{c_3^{(1)}z^2}{c_1^{(1)}}+...= \displaystyle (1+c_1^{(2)}z+c_2^{(2)}z^2+...)^{-1}
\end{align}

This allows the expansion of the denominator of~\eqref{cf1} in terms of another fraction,\vspace{-1mm}
\begin{align}\label{}
\displaystyle c_0+c_1z+c_2z^2+...= c_0+  \displaystyle\frac{c_1^{(0)}z}{1+ \displaystyle \frac{c_1^{(1)}z}{1+c_1^{(2)}z+c_2^{(2)}z^2+...}}
\end{align}

By successively forming the reciprocal series we obtain a $C$-fraction, written in a compact form as,\vspace{-1mm}

\begin{align}\label{cfc}
\displaystyle c_0+c_1z+c_2z^2+...=c_0+{\genfrac{}{}{}{}{c_1^{(0)}z}{1}}   {\genfrac{}{}{0pt}{}{}{+}}   {\genfrac{}{}{}{}{c_1^{(1)}z}{1}}   {\genfrac{}{}{0pt}{}{}{+}}   {\genfrac{}{}{}{}{c_1^{(2)}z}{1}}   {\genfrac{}{}{0pt}{}{}{+\dots}}
\end{align}

By truncating the $C$-fraction in~\eqref{cfc} we obtain its convergents which are rational fractions in $z$. For example the first 4 convergents of~\eqref{cfc} are given as,\vspace{-1mm}
\begin{align}\label{}
\nonumber
&\displaystyle \frac{A_0(z)}{B_0(z)}=c_0, \quad \displaystyle \frac{A_1(s)}{B_1(z)}=c_0+c_1^{(0)}z,\\
&\displaystyle \frac{A_2(z)}{B_2(z)}=\displaystyle \frac{c_0+ (c_0c_1^{(1)}+c_1^{(0)})z}{1+c_1^{(1)}z}, \\
&\displaystyle \frac{A_3(z)}{B_3(z)}=\displaystyle \frac{c_0+ (c_0(c_1^{(1)}+c_1^{(2)})+c_1^{(0)})z+ c_1^{(0)}c_1^{(2)}z^2}{1+(c_1^{(1)}+c_1^{(2)})z} \nonumber
\end{align}

\noindent
where $c_1^{(0)}=c_1$, $c_1^{(1)}=-c_2/c_1$ and $c_1^{(2)}=(c_2^2-c_1c_3)/(c_1c_2)$.


The diagonal Pad\'{e} approximant of degree $M$ of $V(z)$, hereafter appearing frequently in the text, is the (2$M$+1)th convergent of its $C$-fraction representation in~\eqref{cfc},
\vspace{-1mm}
\begin{align}\label{}
\displaystyle \text{PA}[M/M]_V(z)=\frac{A_{2M}(z)}{B_{2M}(z)}
\end{align}

In general a given analytic function can be approximated by PA$[L/M](z)$ where $L$ and $M$ are not necessarily equal,
\vspace{-1mm}
\begin{align}\label{Pade}
\sum_{n=0}^{L+M} c_nz^n=\frac{a_0+a_1z^1+...+a_Lz^L}{b_0+b_1z^1+...+b_Mz^M}+\mathcal{O}(z^{L+M+1})
\end{align}

Setting $b_0=1$, the denominator coefficients $b_1,...,b_M$ are obtained by cross-multiplying~\eqref{Pade}, equating the coefficients of $z^{L+1}$,$z^{L+2}$,...,$z^{L+M}$ to zero and solving the resulting linear system. Next the numerator coefficients $a_0,a_1,...,a_L$ are obtained similarly by considering the coefficients of $z^0$,$z^{1}$,...,$z^{L}$.

\begin{figure}[h!]
    \centering

       \includegraphics[scale=1.55,trim=0cm 0.0cm 0cm 0cm,clip]{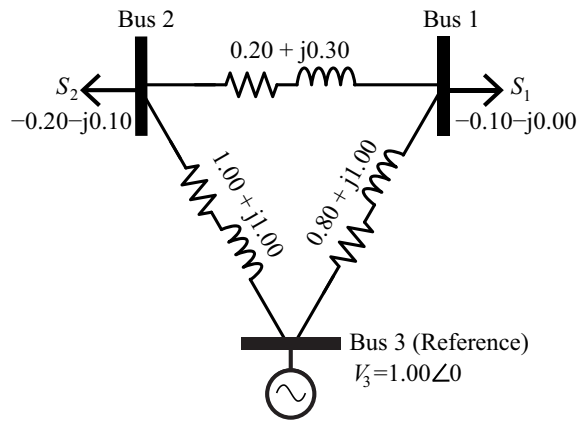}

\caption{3-bus network with no voltage magnitude constraint\vspace{-5mm}} \label{grid1}
\end{figure}

\begin{figure}[h!]
    \centering

       \includegraphics[scale=0.55,trim=0.1cm 0.0cm 0cm 0cm,clip]{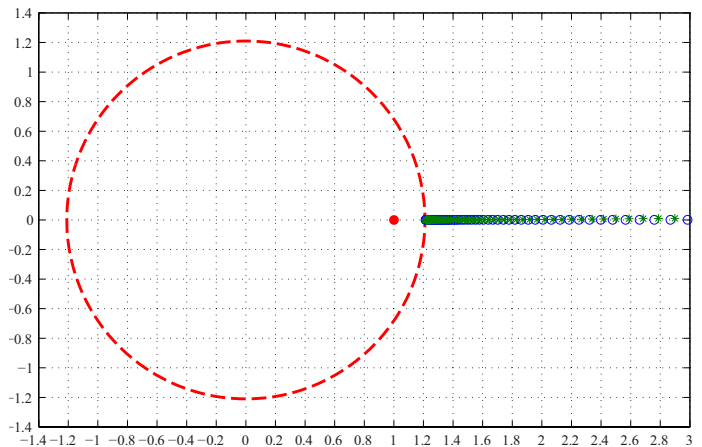}

\caption{Zero-pole distribution of PA[100/100] corresponding to the network of Fig.~\ref{grid1}\vspace{-1mm}} \label{Pade1}
\end{figure}

Now consider the network of Figure~\ref{grid1} where the per-unit values of power flow parameters are shown and the unknown states are voltage phasors $V_1$ and $V_2$. The Taylor series for $V_1(z)$ and $V_2(z)$ are obtained based on~\eqref{HELM2} which are then used to compute the Pad\'{e} coefficients. The concentration of zeros ($o$) and poles ($\ast$) of the diagonal Pad\'{e} approximant, shown in Figure~\ref{Pade1}, defines the closest common branch point of $V_1(z)$ and $V_2(z)$ at $z_\text{b}\hspace{-0mm}\approx\hspace{-0mm}1.2$ which is also given by Fabry's theorem~\cite{Suetin} as $z_\text{b}\hspace{0mm}=\hspace{0mm}\lim_{n\rightarrow\infty}c_n/c_{n+1}$. Note that the branch point is the common limit point of the sequences of zeros and poles (See the discussion of the Stahl's compact set in the Appendix). A closer inspection of the zeros and poles reveals the exact location of the branch point at $1.21510$. This means that if the loading in the network is increased by this factor the new operating point is infeasible. When the loading is increased by $1.21509$, PA[100/100] recovers the solution of the network with active and reactive power mismatches smaller than $10^{-5}$. Here, as it is often the case for the class of problems in~\eqref{p}, i.e. for networks with only PQ buses and a voltage reference, analytic continuation of the germs by rational approximation is unnecessary as all the power series already converge at $z=1$ (highlighted as a red dot) and thus are sufficient for computing $V_1$ and $V_2$. However, Pad\'{e} approximants are much more efficient for evaluating these functions as they converge to a given function at a much higher rate than the original power series does~\cite{Suetin} and can also discover the analytic structure of a given multi-valued function~\cite{ApBuMaSu11}.\vspace{0mm}

\section{Embedding the Voltage Magnitude Constraints in the Complex Plane\protect\footnote{The modelling approaches defined here and in the following section are the improved versions and evolution of our earliest formulation in~\cite{BS15}.}} \label{voltmag}
For a generator node $i \in \mathcal{G}\subset \mathcal{N}$, the real (active) power output, $P_i=\text{Re}(S_i)$, is fixed whereas the imaginary (reactive) power, $Q_i=\text{Im}(S_i)$, is a free parameter which is adjusted so as to fix the magnitude of the voltage phasor $V_i$ at a given setpoint value $M_i$. Extending the general framework of the holomorphic embedding to this case is particularly challenging as here the magnitude of a holomorphic function, $V_i(z)$, is to be held fixed. Such a function has to be constant by the open mapping theorem~\cite{Stein} as the image of $V_i(z)$ in the complex plane is a subset of a circle and thus $V_i$ can no longer be an open map. To resolve this contradiction we define an analytic function $\overline{V}_i(z)=\sum_{n=0}^{\infty} \overline{c}_n^{[i]}z^n $ independently of $V_i(z)$ for $i \in \mathcal{G}$. Note that $\overline{V}_i(z)\neq (V_i(z^*))^*$ and this distinction is the essential concept behind embedding voltage constraints and allows $V_i(z)$ to adopt the value of $V_{r}$ at $z=0$ with its magnitude approaching $M_i$ as $z$ increases. However the challenge remains as how these two independent functions should be related to $M_i$ so that at $z=1$, $\overline{V}_i(z)=(V_i(z^*))^*$ for $i \in \mathcal{G}$.  An approach that enforces $V_i(z)\overline{V}_i(z)=M_i^2$, leaves the possibility that at $z=1$ $V_i(z)\overline{V}_i(z)=M_i^2$  while $\overline{V}_i(z)\neq (V_i(z^*))^*$. In other words the relaxation of $V^*$ into $W$ may not yield a tight solution for the original algebraic equations. To remedy this problem we relax $V^*$ into $W$ for a given generator node $i$ in the following relations,
\vspace{-1mm}
\begin{subequations}\label{genQ}
\begin{align}\label{q}
&\displaystyle P_i+jQ_i= V_i \sum_{k\in \mathcal{N}(i)} V^*_k Y^*_{ik}+ V_i W_i Y^*_{ii} \\
\label{qconj}
&\displaystyle (P_i+jQ_i)^*= W_i \sum_{k\in \mathcal{N}(i)} V_k Y_{ik}+ V_i W_i Y_{ii} \\
&V_iW_i=M_i^2
\end{align}
\end{subequations}

Note that here we have made no assumption that $Q_i$ is real-valued. Since $P$ is real-valued, one can check that $V^*=W$ if and only if $Q_i$ is real-valued, i.e. when $(P_i+jQ_i)^*=P_i-jQ_i$ and by equating the expressions of $Q$ from~\eqref{q} and \eqref{qconj} we reach the following identity relating $V_i$, $W_i$ and $M_i$,
\begin{subequations}
\begin{align}\label{vw}
&\displaystyle W_i (\hspace{-2mm}\sum_{k\in \mathcal{N}(i)} \hspace{-2mm} V_k Y_{ik}+ V_i Y^*_{ii}) \hspace{-0.5mm}= \hspace{-0.5mm}2P_i - M_i^2 Y_{ii}-\hspace{-3.5mm} \sum_{k\in \mathcal{N}(i)}\hspace{-2mm} V_i V_k^* Y_{ik}^*
\end{align}
\end{subequations}
Now if we incorporate \eqref{genQ} into the existing embedding framework of~\eqref{polynomial} and demonstrate that $Q$ is algebraic in $z$ then~\eqref{vw} guarantees that the Taylor series expansion of $Q$ has real coefficients. Consider the following parametrization of the power flow equations augmented by the generator nodes,
\vspace{-1mm}
\begin{subequations}\label{polynomialg}
\begin{align}
\label{poly1n}
&\displaystyle zS^*_i=\sum_{k\in \mathcal{N}[i]}W_i V_k Y_{ik}\hspace{15mm} \forall i\in  \mathcal{N}-\{r\}-\mathcal{G}\\
\label{poly2n}
&\displaystyle zS_i=\sum_{k\in \mathcal{N}[i]} V_i W_k Y^*_{ik} \hspace{15mm} \forall i\in   \mathcal{N}-\{r\}-\mathcal{G}\\
&\displaystyle z(P_i-jQ_i)= \sum_{k\in \mathcal{N}[i]}W_i V_k Y_{ik} \hspace{20mm} \forall i\in  \mathcal{G}\\
&\displaystyle z(P_i+jQ_i)= \sum_{k\in \mathcal{N}[i]} V_i W_k Y^*_{ik} \hspace{20mm} \forall i\in    \mathcal{G}\\
&\displaystyle W_i (\hspace{-2mm}\sum_{k\in \mathcal{N}(i)}\hspace{-2mm} V_k Y_{ik}+ V_i Y^*_{ii})=2P_i - M_i^2 Y_{ii}-\hspace{-3mm} \sum_{k\in \mathcal{N}(i)} \hspace{-2mm}V_i  W_k Y_{ik}^* \nonumber\\
&  \hspace{30mm} \forall i\in \mathcal{G}
\end{align}
\end{subequations}

Notice that in contrast to~\eqref{polynomial} here for each generator node there are three algebraically independent equations and three variables. This parameterization of power flow equations defines {\em generically} an affine algebraic curve in $(z,V_1,V_2,...,W_n,Q_i,...)$. Here again the reduced Gr\"{o}bner basis for a lexicographic order, such as $...>Q_i>z$, gives a bivariate polynomial $f_i(Q_i,z)$ for well-defined problems and thus it follows that $Q_i$ is a multi-valued algebraic function of $z$ and can be represented as a power series expansion. The general principle for defining and analytically continuing a given solution germ can be summed up by reducing the system of~\eqref{polynomialg} in~\eqref{pg}.
\vspace{-1mm}
\begin{subequations}\label{pg}
\begin{align}\label{pg01}
&\displaystyle \frac{zS_i^*}{(V_i(z^*))^*}=\hspace{-2mm}\sum_{k\in \mathcal{N}[i]} V_k(z) Y_{ik} \hspace{16mm} \forall i\in\mathcal{N}-\{r\}\\
\label{pg2}
&\displaystyle \frac{zS_i^*(V,M_i,P_i)}{(V_i(z^*))^*}=\hspace{-2mm}\sum_{k\in \mathcal{N}[i]} V_k(z) Y_{ik} \hspace{18mm} \forall i\in\mathcal{G}
\end{align}
\end{subequations}

Note that a given set of voltage phasors $V$ that satisfies~\eqref{pg} uniquely characterizes a germ to the expanded system of~\eqref{polynomialg} and other variables associated with generator nodes such as $Q$ and $\overline{V}$ are unambiguously defined in relation to this germ. As analytic functions their sole purpose is to facilitate the construction of $S_i^*(V,M_i,P_i)$. Also notice that $W$ and $\overline{V}$ are eliminated in both~\eqref{pg01} and~\eqref{pg2} and this rules out the possibility of ghost solutions where $W_i(z)\neq (V_i(z^*))^*$. With this clarification we propose the following embedding for networks with generator nodes\footnote{There are a number of minor variations to this specific formulation that might be more advantageous from a computational point of view. We chose this formulation as it was simpler to mathematically justify the key condition of $\overline{V}_i(z)=(V_i(z^*))^*$ at $z=1$.},
\begin{subequations}\label{polynomial_G}
\begin{align}
\label{0}
&\displaystyle \frac{zS_i^*}{(V_i(z^*))^*}=\hspace{-2mm}\sum_{k\in \mathcal{N}[i]} V_k(z) Y_{ik} \hspace{16mm} \forall i\in\mathcal{N}-\{r\}\\
\label{1}
&\displaystyle \frac{z(S_i(z^*))^*}{(V_i(z^*))^*}=\hspace{-2mm}\sum_{k\in \mathcal{N}[i]} V_k(z) Y_{ik} \hspace{23mm} \forall i\in\mathcal{G}\\
\label{2}
&\displaystyle \frac{S_i(z)}{V_i(z)}= \overline{V}_i(z) Y^*_{ii} + \hspace{-1mm}\sum_{k\in \mathcal{N}(i)} \hspace{-1mm}(V_k(z^*))^* Y^*_{ik}\hspace{8mm}\forall i\in\mathcal{G} \\
\label{3}
&\overline{V}_i(z)  (\hspace{-2mm}\sum_{k\in \mathcal{N}(i)} \hspace{-2mm} V_k(z) Y_{ik}+ V_i(z) Y^*_{ii})=\nonumber\\
&2P_i - M_i^2 Y_{ii}-\hspace{-3mm} \sum_{k\in \mathcal{N}(i)}\hspace{-2mm} V_i(z) (V_k(z^*))^* Y_{ik}^* \hspace{10mm} \forall i\in\mathcal{G}
\end{align}
\end{subequations}

Here $Q_i(z)$, by construction, has real coefficients and $S_i(z)=P_i+jQ_i(z) \quad \forall i\in\mathcal{G}$. The embedding of~\eqref{polynomial_G} is based on the parametrization introduced in~\eqref{polynomialg} and at $z=1$ sufficiently determines the AC power flow relations in a network with load and generation. Notice the different embedding of~\eqref{1} and~\eqref{2}. The former corresponds to~\eqref{pg2} and in combination with~\eqref{0} is used to develop the germ of the stable branch and ensures that $V_i(z)$ has the trivial solution $V_r$ at $z=0$. The latter constructs $S_i(z)=P_i+jQ_i(z)$ based on the relation that is enforced between $V_i(z)$ and $\overline{V}_i(z)$ in~\eqref{3}. Thanks to this unique construction of $Q_i(z)$ and subsequently $S_i(z)$ one can easily inspect that the combination of~\eqref{1} and~\eqref{2} enforces $\overline{V}_i(z)=(V_i(z^*))^*$ at $z=1$ and as such we succeed in implementing the reduced system of \eqref{pg} while enforcing the voltage magnitude and active power constraints of the generators. The resulting algebraic system is adequately described by the following set of power series relations,\vspace{-1mm}

\begin{figure}
    \centering

       \includegraphics[scale=1.55,trim=0cm 0.0cm 0cm 0cm,clip]{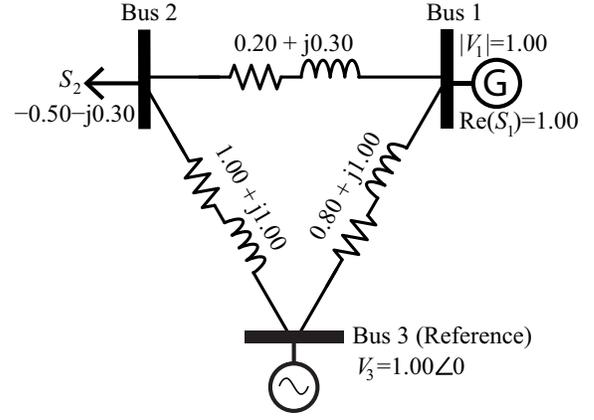}

\caption{3-bus network with voltage control at bus 1\vspace{-3mm}} \label{grid2}
\end{figure}

%

\begin{subequations}\label{HELM1b}
\begin{align}
\label{mainb1}
&zS^*_i\hspace{-1mm}\sum_{n=0}^{\infty}\hspace{-1mm} {d^*_n}^{[i]}z^n\hspace{-0.5mm} =\hspace{-3.0mm}\sum_{k\in \mathcal{N}[i]}\hspace{-2mm}(Y_{ik}\hspace{-1mm}\sum_{n=0}^{\infty}\hspace{-1mm} {c_n}^{[k]}z^n) \hspace{1.5mm} \forall i\in\mathcal{N}\hspace{-1.0mm}-\hspace{-0.50mm}\{r\}\hspace{-0.5mm}-\hspace{-0.5mm}\mathcal{G} \\
\label{mainb2}
&z(\sum_{n=0}^{\infty}\hspace{-0.5mm} {g^*}_n^{[i]}z^n)(\sum_{n=0}^{\infty}\hspace{-1mm} {d^*_n}^{[i]}z^n)\hspace{-0.5mm} =\hspace{-3.0mm}\sum_{k\in \mathcal{N}[i]}\hspace{-2.5mm}(Y_{ik}\hspace{-1.5mm}\sum_{n=0}^{\infty}\hspace{-1mm} {c_n}^{[k]}z^n) \hspace{0.0mm} \forall i\in\mathcal{G}\\
\label{conva}
& \displaystyle (\sum_{n=0}^{\infty} c_n^{[i]}z^n)(\sum_{n=0}^{\infty} d_n^{[i]}z^n)=1 \hspace{17mm} \forall i\in\mathcal{N}-\{r\}\\
\label{convb}
&\displaystyle (\sum_{n=0}^{\infty} \overline{c}_n^{[i]}z^n) (\sum_{k\in \mathcal{N}(i)}\hspace{-2mm}(Y_{ik}\hspace{-1mm}\sum_{n=0}^{\infty}\hspace{-1mm} {c_n}^{[k]}z^n)+ Y_{ii}^*\hspace{-1mm}\sum_{n=0}^{\infty}\hspace{-1mm} {c_n}^{[i]}z^n)= \nonumber\\
&M_i^2Y_{ii}-2P_i+\hspace{-3mm}\sum_{k\in \mathcal{N}(i)}\hspace{-2mm}(Y_{ik}^*\hspace{-1mm}\sum_{n=0}^{\infty}\hspace{-1mm} {c^*_n}^{[k]}z^n)(\sum_{n=0}^{\infty}\hspace{-1mm} {c_n}^{[i]}z^n) \forall i\in\mathcal{G}\\
\label{convc}
&(\sum_{n=0}^{\infty} g_n^{[i]}z^n)(\sum_{n=0}^{\infty} d_n^{[i]}z^n) =\nonumber\\
&Y^*_{ii}\hspace{-0.5mm}\sum_{n=0}^{\infty} \overline{c}_n^{[i]}z^n+\hspace{-3mm}\sum_{k\in \mathcal{N}(i)}\hspace{-3mm}(Y^*_{ik}\hspace{-0.5mm}\sum_{n=0}^{\infty}\hspace{-0.5mm} {c^*_n}^{[k]}z^n) \hspace{17mm} \forall i\in\mathcal{G}
\end{align}
\end{subequations}

The key coefficients $c_n^{[i]}$ ($\forall i\in\mathcal{N}\hspace{-1.0mm}-\hspace{-0.50mm}\{r\}$) are progressively obtained by differentiating \eqref{mainb1} and \eqref{mainb2} with respect to $z$, evaluating at $z=0$ and solving the resulting linear system which itself requires the prior knowledge of $d_m^{[i]}$, $\overline{c}_m^{[i]}$ and $g_m^{[i]}$ for $m=1,...,n-1$. These coefficients are already obtained at previous stages. Once this linear system of size $|\mathcal{N}|-1$ is solved for $c_n^{[i]}$ ($\forall i\in\mathcal{N}\hspace{-1.0mm}-\hspace{-0.50mm}\{r\}$), $\overline{c}_n^{[i]}$ ($\forall i\in\mathcal{G}$) are obtained from~\eqref{convb}. Then we can compute $\overline{g}_n^{[i]}$ ($\forall i\in\mathcal{G}$) from~\eqref{convc} and repeat the previous steps to obtain the next set of coefficients. Notice that by construction $g_0^{[i]}=P_i+jq_0^{[i]}$ and $g_n^{[i]}=jq_n^{[i]}$ for $n\ge1$ where $q^{[i]}$, the coefficients of $Q_i(z)$, are real-valued. It is straightforward to enforce the reactive limit of a generator node $i$ which changes $(S_i(z^*))^*$ to $S_i^*$ in \eqref{1} and removes the corresponding equations in~\eqref{2}-\eqref{3}. Accordingly the corresponding power series on the left side of \eqref{mainb2} is replaced by~\eqref{mainb1} and the ones in~\eqref{convb} and~\eqref{convc} are eliminated.

\begin{figure*}
\begin{align}\label{m1}
\footnotesize
\arraycolsep=2.5pt
\medmuskip = 0.0mu
\scalefont{1}{
\begin{pmatrix*}[r]
0 & 0 & 0 & 0 & 0 & 0 & 0 \\
 0 & 0.00251-0.00171 i & 0.06219-0.04894 i & 0.07360-0.19166 i & 0.10368+0.00834 i & -0.04226-0.04332 i & \hspace{2mm}0.00019+0.00524 i \\
 0.00106-0.00018 i & 0.01831-0.02968 i & -0.05020-0.56373 i & -0.46670-0.47703 i & 0.38542-0.41374 i & -0.22562+0.08679 i & 0.02509-0.00623 i \\
 0.00065-0.00842 i & -0.03951-0.10721 i & -1.53037+0.00275 i & -0.93086+1.70712 i & -0.68418-1.05204 i & 0.01889+0.58108 i & 0.04075-0.04565 i \\
 -0.03691+0.00834 i & -0.19620+0.26933 i & 0.72469+3.04445 i & 3.55535-1.19261 i & -0.91634+0.33716 i & 0.33003+0.12123 i & 0.01180-0.06678 i \\
 0.03275+0.16840 i & 0.61071+0.63059 i & 1.58947-2.87746 i & -2.37011-1.93190 i & 0.11278+0.83042 i & 0.08156-0.12479 i & 0 \\
 0.27347-0.00447 i & -0.48693-1.03924 i & -1.37499+0.56371 i & 0.15033+0.76564 i & 0.00650-0.14824 i & 0 & 0\\
 -0.27099-0.16367 i & -0.22579+0.38863 i & 0.29999+0.04936 i & 0 & 0 & 0 & 0
\end{pmatrix*}
}
\end{align}

\begin{align}\label{mb}
\footnotesize
\arraycolsep=2.5pt
\medmuskip = 0.0mu 
\scalefont{1}{
\begin{pmatrix*}[r]
0 & 0 & 0 & 0 & 0 & 0 & 0 \\
0 & 0.00058+0.00030 i & 0.01500+0.00427 i & 0.03594-0.05669 i & -0.02112+0.10596 i & 0.02158-0.06639 i & -0.00479+0.00533 i \\
 0.00012+0.00009 i & 0.00681-0.00255 i & 0.09960-0.01779 i & 0.14498-0.10685 i & 0.07458+0.50024 i & -0.00671-0.34180 i & 0.03068+0.01416 i \\
 0.00091-0.00078 i & -0.00130-0.00754 i & 0.19685+0.00828 i & 0.39108+0.66032 i & -0.20047+0.44572 i & 0.31730-0.43239 i & 0.06940-0.00611 i \\
 -0.00452-0.00286 i & -0.01718+0.11334 i & 0.35001+0.33909 i & -1.13221+0.70117 i & 0.54878+0.53354 i & 0.45355-0.23719 i & 0.02803-0.02035 i \\
 -0.01357+0.02012 i & 0.18759+0.15906 i & -1.14552-0.28566 i & -0.26170+0.94492 i & 0.67023+0.07099 i & 0.03391-0.03513 i & 0 \\
 0.02779+0.02690 i & -0.24357-0.30435 i & -0.37291+0.24887 i & 0.25444+0.19442 i & 0.01547-0.01417 i & 0 & 0\\
 -0.01073-0.04347 i & -0.06144+0.01628 i & 0.02506+0.03494 i & 0 & 0 & 0 & 0
\end{pmatrix*}
}
\end{align}
\vspace{-2mm}
\end{figure*}

Now consider the modified network of Figure~\ref{grid2} where bus 1 has a generator that regulates its voltage magnitude at $1.00$ and generates $P_1=1.00$. Consider the embedding of~\eqref{polynomialgn}.\vspace{-3mm}

\begin{subequations}\label{polynomialgn}
\begin{align}
\label{e2a}
&\displaystyle z(P_1+jQ_1)= V_1 (W_1 Y^*_{11}+W_2 Y^*_{12}+V_3^* Y^*_{13})\\
\label{e2b}
&\displaystyle z(P_1-jQ_1)= W_1 (V_1 Y_{11}+V_2 Y_{12}+V_3 Y_{13})\\
\label{e2c}
&\displaystyle P_1+jQ_1= V_1 (\overline{V}_1 Y^*_{11}+ W_2 Y^*_{12}+V_3^* Y^*_{13})\\
\label{e2d}
&\displaystyle \overline{V}_1 (V_2 Y_{12}+V_3 Y_{13}+ V_1 Y^*_{11})=\nonumber\\
&\displaystyle 2P_1 - M_1^2 Y_{11}-V_1(W_2 Y_{12}^*+V_3^* Y^*_{13})\\
\label{e1a}
&\displaystyle zS_2=V_2 (W_1 Y^*_{12}+W_2 Y^*_{22}+V_3^* Y^*_{23})\\
\label{e1b}
&\displaystyle zS^*_2=W_2 (V_1 Y_{12}+V_2 Y_{22}+V_3 Y_{23})
\end{align}
\end{subequations}

Here $z$ is the embedding parameter and the unknown states are $(V_1, W_1, V_2, W_2,\overline{V}_1,Q_1)$. All other quantities are parameters of the power flow problem.
Notice that~\eqref{e2a}-\eqref{e2d} represent the embedding of the power flow relations of bus 1 (PV) with~\eqref{e2c}-\eqref{e2d} modeling the action of the generator AVR in the complex plane enforcing $|V_1|=|\overline{V}_1|=M_1$ at $z=1$. Equations~\eqref{e1a}-\eqref{e1b} correspond to the power flow relations of bus 2 (PQ).

From a geometric point of view, the algebraically independent equations of~\eqref{polynomialgn} define an affine algebraic variety of dimension 1, i.e. an algebraic curve in $(z,V_1, W_1, V_2, W_2,\overline{V}_1,Q_1)$. There is a polynomial ideal $I \subset \mathbb{C} [ z,V_1, W_1, V_2, W_2,\overline{V}_1,Q_1 ]$ corresponding to this algebraic variety and the equations of~\eqref{polynomialgn} are only one basis, among many different bases, that generate $I$~\cite{Cox}. These equations can be the starting basis for finding the corresponding reduced Gr\"{o}bner basis of $I$ with the unique property that for a given lexicographic (lex) order, such as $...>V_1>z$, the last element of this special basis is a bivariate polynomial $f(V_1,z)$. The elements of the reduced Gr\"{o}bner basis have a triangular structure in terms of the appearance of the unknown states and $z$ and, as such, the algorithms for obtaining this special basis are the nonlinear generalization of the Gaussian elimination in linear algebra.

To highlight the significance of this analysis for voltage collapse studies let us compute $f(V_2,z)$ first for the set of power flow parameters shown in Figure~\ref{grid2} and next for when the power flow reaches its feasibility boundary at $P_1=2.6785$. The coefficient matrices of these two bivariate polynomials which have the form $\sum_{i=0}^{6} \sum_{k=0}^{6} a_{ik} V_2^iz^k$ are shown in~\eqref{m1} and~\eqref{mb} respectively. Note that the rows of the matrices correspond to $(1,V_2,V_2^2,...,V_2^6)$ and the columns correspond to $(1,z,z^2,...,z^6)$. So the element in the $i$th row and the $k$th column is the coefficient of $V^{i-1}z^{k-1}$. We can recover all solutions of the power flow by simply setting $z=1$ and solving for the roots of the resulting univariate polynomial in $V_2$. This will include all valid solutions of $V_2$ as well as solutions where $V_2\neq W_2^*$. By evaluating the next basis element at $z=1$ and a given solution of $V_2$ we can obtain the numerical value of the next unknown in the lex order, for example $V_1$ if the lex order was chosen as $...>V_1>V_2>z$. This value of $V_1$ corresponds to that particular solution of $V_2$. However this is not the focus of our paper. The critical points of the projection of these algebraic curves onto $\mathbb{\overline{C}}$, the {\em extended complex plane}, are where $\partial{f_i(V_i,z)}/\partial{V_i}=0$. A subset of these so-called $z$-critical points~\cite{Geom} for the two cases of $P_1=1.0000$ and $P_1=2.6785$ are shown in Table~\ref{BPnts}. These points, as explained later, are the branch points of power flow variables, considered as analytic functions of a single complex variable, i.e. $z$ and play an important role in the analytic continuation of the germ from a trivial solution to the actual power flow solution and contain vital information on the voltage stability margin of the operating point. However there is absolutely no need to compute the reduced Gr\"{o}bner basis in order to locate these branch points in $\mathbb{\overline{C}}$. In fact computation of these basis is exponential space complete~\cite{Mayr} and requires time that is at least exponential in the number of solutions of the polynomial system. Hence there is little prospect in the near future for applying the concept of reduced Gr\"{o}bner basis to industrial power flow studies.

\begin{table}
\caption{} \label{BPnts}
\centering
\medmuskip = 0.0mu
\renewcommand{\arraystretch}{1.0}
\begin{tabular} {|c||l|l|} \hline 
 Branch        & $\hspace{2mm}P_1=1.0000$ & $\hspace{2mm}P_1=2.6785$ \\
 Points     &      & \hspace{2mm} Bifurcation \\\hline
$z_{\text{B}}$&\hspace{2.35mm}2.5742&\hspace{2.35mm}1.0000\\\hline
$z_1$&$-0.8496$&$-0.2266$\\\hline
$z_2$&$-0.3672+0.6263i$&$-0.6170+1.1219i$\\\hline
$z_3$&$-0.3672-0.6263i$&$-0.6170-1.1219i$\\\hline
$z_4$&$-0.4644+1.2672i$&\hspace{2.35mm}$0.7611+0.9593i$\\\hline
$z_5$&$-0.4644-1.2672i$&\hspace{2.35mm}$0.7611-0.9593i$\\\hline
$z_6$&&\hspace{2.35mm}$0.9638+0.2129i$\\\hline
$z_7$&&\hspace{2.35mm}$0.9638-0.2129i$\\\hline
\end{tabular}
\vspace{-5mm}
\end{table}

\begin{figure*}
        \centering
    \subfloat[The Stahl's compact set when the operating point is far from the power flow feasibility boundary ($P_2=1.0000$).\label{Pade2}]{                \includegraphics  [scale=0.95,trim=0.0cm 0.0cm 0.0cm 0.0cm,clip]{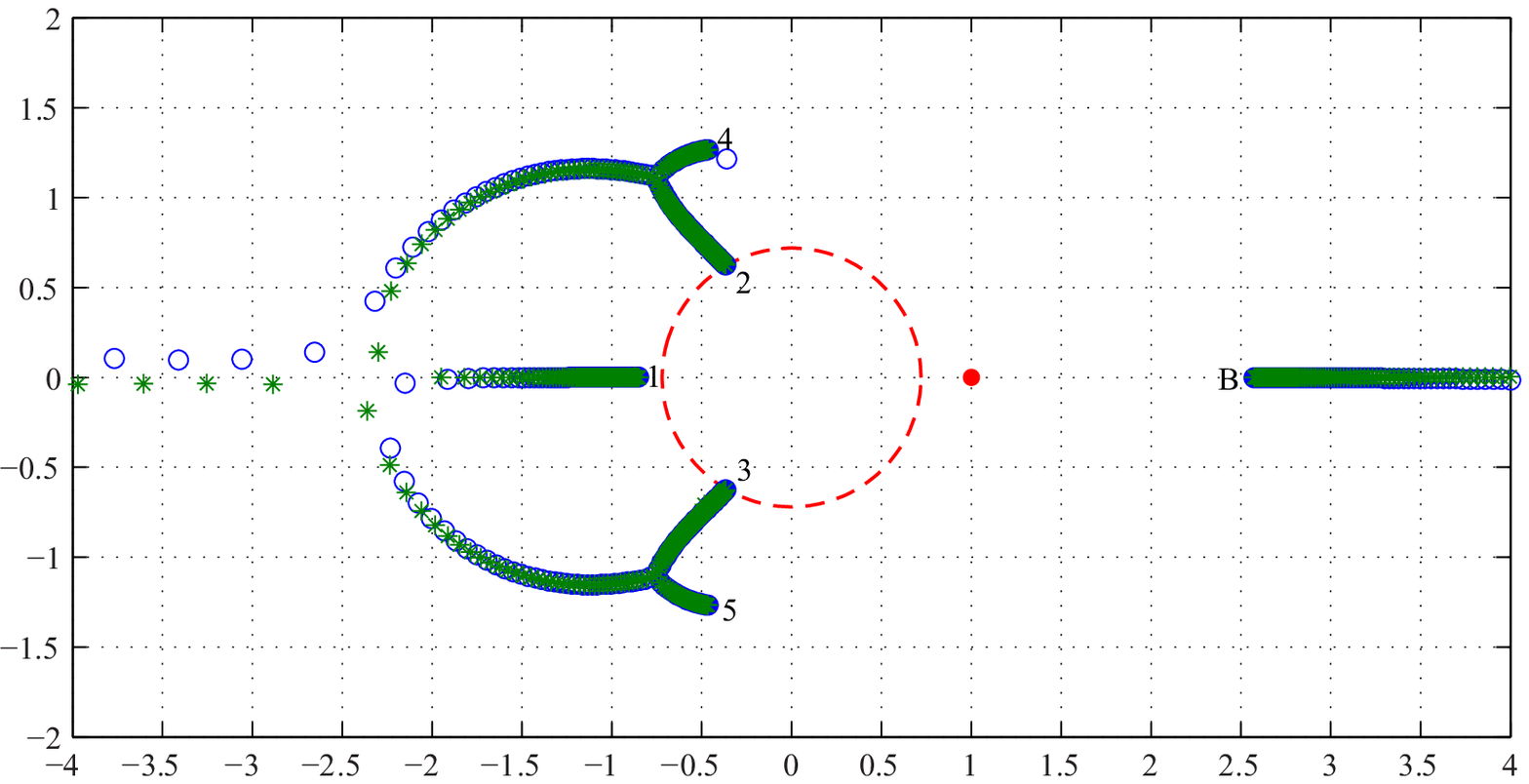}}
           \\
        \subfloat[Transformation of the Stahl's compact set when the operating point is precisely on the feasibility boundary ($P_2=2.6785$)\label{Pade_bf}]{
        \includegraphics [scale=0.95,trim=0.0cm 0.cm 0.0cm 0.0cm,clip]{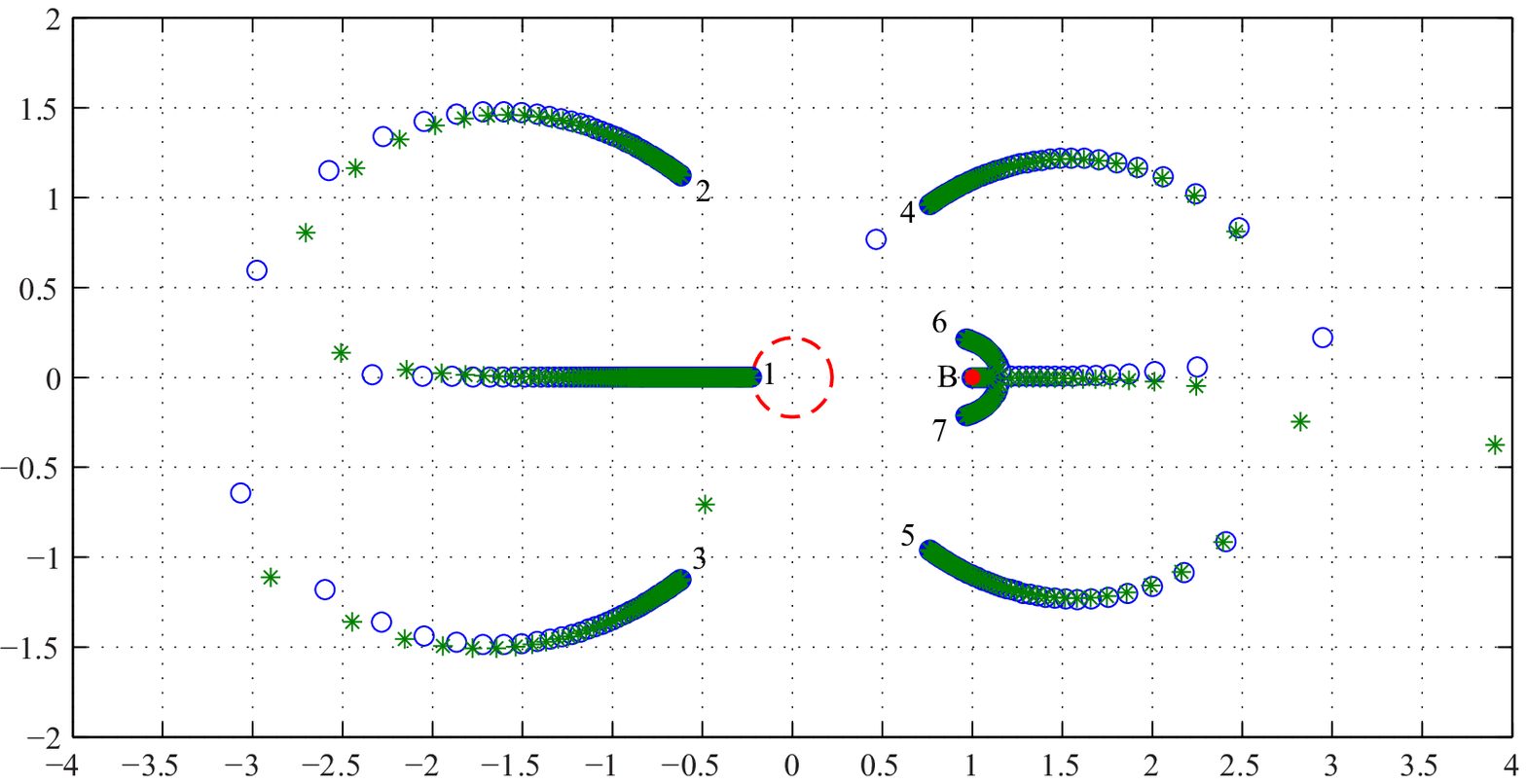}}
         \caption{Zero-pole distributions of PA[1000/1000] depicting the Stahl's compact set, i.e. the analytic structure and the common branch points of $V_1(z)$, $\overline{V}_1(z)$, $V_2(z)$ and $Q_2(z)$ (corresponding to the network of Fig.~\ref{grid2}).\vspace{0mm}}\label{Pade}
       \includegraphics[scale=0.95,trim=0.0cm 0.cm 0.0cm 0.0cm,clip]{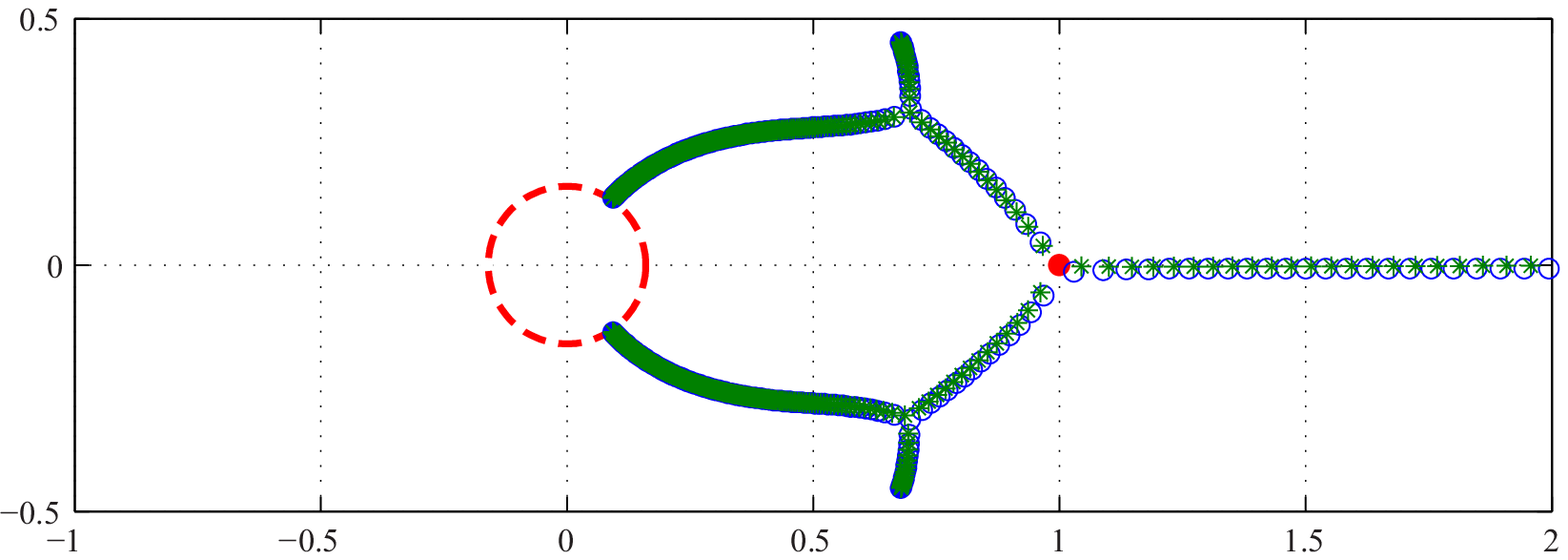}

\caption{Zero-pole distribution of PA[1000/1000] depicting the Chebotarev's point reaching $z=1$ at bifurcation point (corresponding to Fig.~\ref{grid2} and the approach defined by~\eqref{sub})} \label{dymarsky}
\end{figure*}

Instead we develop the germs of the unknown states $(V_1,V_2,\overline{V}_1,Q_1)$, according to~\eqref{HELM1b}. Figure~\ref{Pade} shows the zero-pole distribution of the diagonal Pad\'{e} approximant for $V_1(z)$ forming the Stahl's compact sets corresponding to the network of Figure~\ref{grid2} for $P_1=1.0000$ (Figure~\ref{Pade2}) and $P_1=2.6785$ (Figure~\ref{Pade_bf}). Notice that the analytic arcs (branch cuts) of voltage phasors are highlighted by the distribution of zeros ($o$) and poles ($\ast$) of the truncated $C$-fraction, i.e. Pad\'{e} approximant. The points of infinite density of the zeros and poles are exactly (within 5 decimal digits for PA[1000/1000]) the branch points (cf. Table~\ref{BPnts}) of $V_1(z)$ which, in this case, are common with $\overline{V}_1(z)$, $V_2(z)$ and $Q_1(z)$. These branch points are a subset the $z$-critical points of the algebraic curve that we computed earlier as the last element of reduced Gr\"{o}bner basis of the polynomial system of~\eqref{polynomialgn}. An analytic arc emanates from each branch point and culminates in a different branch point or in a Chebotarev's point of the Stahl's compact set. The region of convergence is a disk bounded by the closest branch points, a pair of complex conjugate points at $z_\text{b}=-0.3672\pm j0.6263$. In contrast to the previous case where the region of convergence of the series contained $z=1$, here, the concept of analytic continuation by Pad\'{e} approximants is elegantly illustrated. Since $\lim_{n\rightarrow\infty}|c_{n+1}|/|c_{n}|\approx 1.4$, the coefficients tend to explode rapidly.  Without Pad\'{e} approximants based on these otherwise useless coefficients, it is impossible to recover the power flow solution.

Figure~\ref{Pade_bf} shows the transformation of the Stahl's compact set as the power flow solution reaches the feasibility boundary at $P_1=2.6785$. The branch point on the positive real axis has now moved to $z=1$. Since past the branch point, there is a non-trivial monodromy, examining the PA solutions, as the degree of the diagonal Pad\'{e} approximants is increased, reveals whether the power flow problem has a stable solution or not. This procedure is shown concretely in sections V and VI. The location of this branch point can also serve as a proximity index to the feasibility boundary where the saddle-node bifurcation, i.e. loss of structural stability, occurs. It is worth mentioning that the power flow is still feasible at $P_1=2.6785$ and the PA method recovers the solution whereas Newton-Raphson method fails to converge for $P_1 \ge 2.6750$. 

\section{An Alternative Approach to Embedding the Voltage Magnitude Constraints in the Complex Plane}
In Section~\ref{voltmag} we essentially constructed $\displaystyle S_i^*(V,M_i,P_i)$ ($\forall i\in\mathcal{G}$) to be incorporated into~\eqref{p}. This involved, at each stage, a matrix-vector multiplication to obtain the coefficients of voltage phasors $V$ which were subsequently used to obtain the coefficients of $\overline{V}_i$ and later the {\em real-valued} coefficients of $Q_i$ ($\forall i\in\mathcal{G}$). The matrix dimension for developing the power series expansion of voltage phasors was the same as the number of load and generator nodes, i.e. $|\mathcal{N}|-1$. An alternative approach is to develop the power series expansion of $V$, $\overline{V}$ and $Q$ simultaneously. For $Q$ to have real-valued coefficients, the auxiliary variables $\overline{V}$ are extended to load nodes, in contrast to the previous approach which $\overline{V}$ was exclusively defined for generator nodes. The embedded equations take the following form,

\begin{subequations}\label{alternative}
\begin{align}
\label{0a}
&\displaystyle \frac{zS_i^*}{(V_i(z^*))^*}=\sum_{k\in \mathcal{N}[i]} V_k(z) Y_{ik} \hspace{14mm} \forall i\in\mathcal{N}-\{r\}\\
\label{1a}
&\displaystyle \frac{z(S_i(z^*))^*}{(V_i(z^*))^*}=\sum_{k\in \mathcal{N}[i]} V_k(z) Y_{ik} \hspace{23mm} \forall i\in\mathcal{G}\\
\label{0ab}
&\displaystyle \frac{zS_i}{(\overline{V}_i(z^*))^*}=\sum_{k\in \mathcal{N}[i]} \overline{V}_k(z) Y^*_{ik} \hspace{14mm} \forall i\in\mathcal{N}-\{r\}\\
\label{1ab}
&\displaystyle \frac{zS_i(z)}{(\overline{V}_i(z^*))^*}=\sum_{k\in \mathcal{N}[i]} \overline{V}_k(z) Y^*_{ik} \hspace{23mm} \forall i\in\mathcal{G}\\
\label{2a}
&V_i(z)\overline{V}_i(z)=V_r^2+z(M_i^2-V_r^2) \hspace{18mm} \forall i\in\mathcal{G}
\end{align}
\end{subequations}

\noindent
where $S_i(z)=P_i+jQ(z)$ and $(S_i(z^*))^*=P_i-jQ(z)$ in~\eqref{1a} and~\eqref{1ab} with $Q_i(z)$ coefficients being \emph{real-valued} a fact that follows from the symmetry of embedded equations~\eqref{0a} through~\eqref{1ab} in combination with~\eqref{2a} which enforces a complex-conjugate relationship\footnote{This condition can be exploited to enhance the computational performance of this approach and we will explain this in a future publication on the computational aspects of different embedding approaches.} between the coefficients of $\overline{V}_i(z)$ and those of $V_i(z)$. Note that at $z=0$ $V_i(z)=\overline{V}_i(z)=V_r$ and this characterizes the unstressed (zero-current) state of the network. The resulting algebraic system is adequately described by the following set of power series relations,\vspace{-3mm}

\begin{subequations}\label{HELM1ab}
\begin{align}
\label{mainbb1}
&\displaystyle zS^*_i\hspace{-1mm}\sum_{n=0}^{\infty}\hspace{-1mm} {d^*_n}^{[i]}z^n\hspace{-0.5mm} =\hspace{-3.0mm}\sum_{k\in \mathcal{N}[i]}\hspace{-2mm}(Y_{ik}\hspace{-1mm}\sum_{n=0}^{\infty}\hspace{-1mm} {c_n}^{[k]}z^n) \hspace{1.5mm} \forall i\in\mathcal{N}\hspace{-1.0mm}-\hspace{-0.50mm}\{r\}\hspace{-0.5mm}-\hspace{-0.5mm}\mathcal{G} \\
\label{mainbb2}
&\displaystyle z(\sum_{n=0}^{\infty}\hspace{-0.5mm} {g^*}_n^{[i]}z^n)(\sum_{n=0}^{\infty}\hspace{-1mm} {d^*_n}^{[i]}z^n)\hspace{-0.5mm} =\hspace{-3.0mm}\sum_{k\in \mathcal{N}[i]}\hspace{-2.5mm}(Y_{ik}\hspace{-1.5mm}\sum_{n=0}^{\infty}\hspace{-1mm} {c_n}^{[k]}z^n) \hspace{0.0mm} \forall i\in\mathcal{G}\\
\label{mainab1}
&\displaystyle zS_i\hspace{-1mm}\sum_{n=0}^{\infty}\hspace{-1mm} {{\overline{d}}^*_n}^{[i]}z^n\hspace{-0.5mm} =\hspace{-3.0mm}\sum_{k\in \mathcal{N}[i]}\hspace{-2mm}(Y^*_{ik}\hspace{-1mm}\sum_{n=0}^{\infty}\hspace{-1mm} {\overline{c}_n}^{[k]}z^n) \hspace{1.5mm} \forall i\in\mathcal{N}\hspace{-1.0mm}-\hspace{-0.50mm}\{r\}\hspace{-0.5mm}-\hspace{-0.5mm}\mathcal{G} \\
\label{mainab2}
&\displaystyle z(\sum_{n=0}^{\infty}\hspace{-0.5mm} g_n^{[i]}z^n)(\sum_{n=0}^{\infty}\hspace{-1mm} {\overline{d}^*_n}^{[i]}z^n)\hspace{-0.5mm} =\hspace{-3.0mm}\sum_{k\in \mathcal{N}[i]}\hspace{-2.5mm}(Y^*_{ik}\hspace{-1.5mm}\sum_{n=0}^{\infty}\hspace{-1mm} {\overline{c}_n}^{[k]}z^n) \hspace{1.0mm} \forall i\in\mathcal{G}\\
\label{voltm}
&\displaystyle(\sum_{n=0}^{\infty}\hspace{-1mm} {c_n}^{[i]}z^n)(\sum_{n=0}^{\infty}\hspace{-1mm} {\overline{c}_n}^{[i]}z^n)\hspace{-1mm}=\hspace{-1mm}V_r^2+z(M_i^2-V_r^2) \hspace{1.0mm} \forall i\in\mathcal{G}\\
\label{convaa}
& \displaystyle (\sum_{n=0}^{\infty} c_n^{[i]}z^n)(\sum_{n=0}^{\infty} d_n^{[i]}z^n)=1 \hspace{17mm} \forall i\in\mathcal{N}-\{r\}\\
\label{convba}
& \displaystyle (\sum_{n=0}^{\infty} \overline{c}_n^{[i]}z^n)(\sum_{n=0}^{\infty} \overline{d}_n^{[i]}z^n)=1 \hspace{17mm} \forall i\in\mathcal{N}-\{r\}
\end{align}
\end{subequations}

The coefficients $c_n^{[i]}$ and $\overline{c}_n^{[i]}$ ($\forall i\in\mathcal{N}\hspace{-1.0mm}-\hspace{-0.50mm}\{r\}$) and $g_n^{[i]}$ ($\forall i\in\mathcal{G}$) are progressively obtained by differentiating \eqref{mainbb1} through \eqref{voltm} with respect to $z$, evaluating at $z=0$ and solving the resulting linear system which itself requires the prior knowledge of $c_m^{[i]}$, $d_m^{[i]}$, $\overline{c}_m^{[i]}$, $\overline{d}_m^{[i]}$ and $g_m^{[i]}$ for $m=1,...,n-1$, already obtained at previous stages. This involves matrix-vector multiplication with the size of matrix being $2|\mathcal{N}|+|\mathcal{G}|-2$ (Contrast with $|\mathcal{N}|-1$ in Section~\ref{voltmag}). Notice that here, similar to the previous approach defined in~\eqref{HELM1b}, $g_0^{[i]}=P_i+jq_0^{[i]}$ and $g_n^{[i]}=jq_n^{[i]}$ for $n\ge1$ where $q^{[i]}$, the coefficients of $Q_i(z)$, are real-valued.

The distinction between this approach and the one proposed earlier in Section III is primarily related to the required degree of Pad\'{e} approximants to achieve a certain level of solution accuracy. We have noticed in all cases examined so far that with the exception of a few this alternative approach requires lower degrees of Pad\'{e} approximants. However, the advantage of the first method is the smaller dimension of the matrix used in developing the voltage phasor power series. This dimension is expected to be critical for the computational performance of the embedding method on extremely large networks. The Pad\'{e} degree requirement of each of these two approaches will be examined in the context of medium and large-scale networks in Part III of this paper.

\section{A Problem with a previously proposed approach}

Previously, a different approach to modeling PV (generator) nodes was proposed that eliminates reactive power by manipulating the power flow equations of the generator node in the following way~\cite{Subramanian},
\begin{subequations}
\begin{align}\label{sub}
&M_i^2 \sum_{k\in \mathcal{N}[i]}Y_{ik} V_k= 2P_iV_i- V_i^2\sum_{k\in \mathcal{N}[i]}Y^*_{ik} V^*_k
\end{align}
\end{subequations}

This formulation can be incorporated into the general framework of the holomorphic embedding for developing the germ solution. However there are two major issues with this approach. First~\eqref{sub} does not adequately constrain the voltage magnitude and one can check that the analytically continued power series of the network of Figure~\ref{grid2}, obtained based on the specific embedded form of~\eqref{sub} as presented in equation (20) of reference~\cite{Subramanian}, satisfies neither the voltage magnitude nor the active power constraints. So there is a need to consider $\overline{V}_i(z)$. However, since reactive power is not considered, it is difficult to see how a relation similar to~\eqref{3} can be enforced in this case to guarantee $\overline{V}_i(z)=(V_i(z^*))^*$ at $z=1$. Even under the assumption that a stronger relation is enforced between $\overline{V}_i(z)$, $V_i(z)$ and $M_i$, for example $\overline{V}_i(z)V_i(z)=M_i^2$ or $\overline{V}_i(z)V_i(z)=(1-z)V_{r}^2+zM_i^2$ in the case of the network of Figure~\ref{grid2}, there arises an even more fundamental problem. In Figure~\ref{Pade_bf} we showed that the branch point on the real-axis reaches $z=1$ as the system experiences saddle-node bifurcation. In the approach based on~\eqref{sub} it is the Chebotarev's point that reaches $z=1$ (see Figure~\ref{dymarsky}). Chebotarev's point is a 0-density point for the equilibrium measure that is concentrated on the Stahl's compact set.
In the generic case, from each Chebotarev's point exactly three analytic arcs
emanate and each arc culminates at either another Chebotarev's point or a branch point
of $f$. In contrast, from each branch point of $f$ only a single analytic arc emanates and culminates at either a Chebotarev's point or
another branch point. Thus the local structure of the Stahl's compact set at a branch point is different from its local structure at a
Chebotarev's point.

The rate of convergence of Pad\'{e} approximants at a given point $z$ declines as that point gets closer to the Stahl's compact set and this can be explained in terms of the value of the Green's function $g_S(z,0)$ (see the Appendix). However this is not the unique reason for the rate of convergence to decline. The structure of the Stahl's compact set can also cause the rate of convergence to decline. Notice the peculiar form of the analytic arcs in Figure~\ref{dymarsky} forming a pair of pincers that encompass the segment of the real axis from $z\approx0.1$ to $z=1$. In particular notice the small gap at the opening of the pincers. As this gap shrinks, the value of the Green's function for the space contained within the pincers gets smaller. No matter how large the space inside the pincers is, the small gap at the opening of the pincers causes the rate of convergence of Pad\'{e} approximants to suffer {\em drastically} and this makes the effective analytic continuation of the germ impossible when the Chebotarev's point is close to $z=1$. This has implications for power flow studies near the feasibility boundary where embedding the power flow in the complex plane is most needed as other methods tend to fail. By contrasting these two approaches we intend to highlight the significance of the correct embedding approach both for efficiently solving the power flow and also for obtaining a reliable proximity index to power flow infeasibility and voltage collapse based on the zero-pole distribution of the rational approximants.

\section{Exponential and ZIP load models}

In real systems loads are voltage dependent and their representation as constant parameters may render the power flow analysis grossly inaccurate.  In this section we describe how the exponential load model and its special case ZIP load model can be incorporated into the framework of embedding the power flow in the complex plane. Exponential load model is described as, 
\begin{subequations}
\begin{align}\label{EL0}
\displaystyle P=|V|^a P_0\\
Q=|V|^b Q_0
\end{align}
\end{subequations}

\noindent
where $a$ and $b$ are rational constants and each can be expressed as a fraction of two relatively prime integers $(m,n)$. Consider the following relations assuming $a=b$ (we later relax this constraint),
\begin{subequations}\label{EL}
\begin{align}
\displaystyle VI^*=|V|^{\frac{m}{n}}S_0\\
V^*I=|V|^{\frac{m}{n}}S_0^*
\end{align}
\end{subequations}

\noindent
which are equivalent to,
\begin{subequations}\label{EL2}
\begin{align}
\displaystyle V^{mn}(I^*)^{mn}=|V|^{m^2}S_0^{mn}\\
|V|^{2mn}I^{mn}=|V|^{m^2}V^{mn}(S_0^*)^{mn}
\end{align}
\end{subequations}

From~\eqref{EL} we have $\displaystyle II^*=S_0S_0^*(VV^*)^{\frac{m-n}{n}}$ which transforms~\eqref{EL2} into a purely phasor relation between the current, voltage and apparent power of an exponential load given as,

\begin{align}\label{EL3}
\displaystyle I^{2n}=\frac{V^m (S_0^*)^{2n}}{(V^*)^{2n-m}}
\end{align}

Now suppose $a=m/n$ and $b=r/s$ in~\eqref{EL0} then we can consider $I$ as the sum of $I_p$ and $I_q$ given by,
\begin{subequations}\label{ELpq}
\begin{align}
\displaystyle I_p^{2n}=\displaystyle \frac{V^m P_0^{2n}}{(V^*)^{2n-m}}\\
\displaystyle I_q^{2s}=\frac{V^r (-jQ_0)^{2s}}{(V^*)^{2s-r}}
\end{align}
\end{subequations}

One can check that the case of $m=2$, $n=1$ in~\eqref{EL3} corresponds to constant-impedance load and the case of $m=1$, $n=1$ models constant-current load. By putting $m=0$ in~\eqref{EL3} we obtain the familiar constant-power load model. Hence the ZIP load model can be expressed as a relation between each current component satisfying~\eqref{EL3} for its corresponding $(m,n)$ and the net current being the sum of each component current similar to $I=I_p+I_q$ in~\eqref{ELpq}.

Now that we have established the algebraic relations between the phasor quantities of various components of voltage-dependent load model we can consider each component current as an algebraic function in $z$. The general principle is similar to the case of voltage magnitude constraint where we develop the germ of voltage phasors by constructing $S_i^*(V,M_i,P_i)$ for a given generator node $i$. Here we construct $I_{i}(V_i,S_i,m,n)$ which is again unambiguously defined in relation to the germ of the stable solution. The corresponding equation that should be added to~\eqref{pg} is as follows,

\begin{figure*}
        \centering

    \subfloat[Zero-pole concentration forms the Stahl's compact set highlighting the common branch points of $V_1(z)$, $\overline{V}_1(z)$, $S_1(z)$ and $V_2(z)$ ($P_1=1.00$).\label{ZIP}]{                \includegraphics  [scale=1.73,trim=0.0cm 0.0cm 0.0cm 0.0cm,clip]{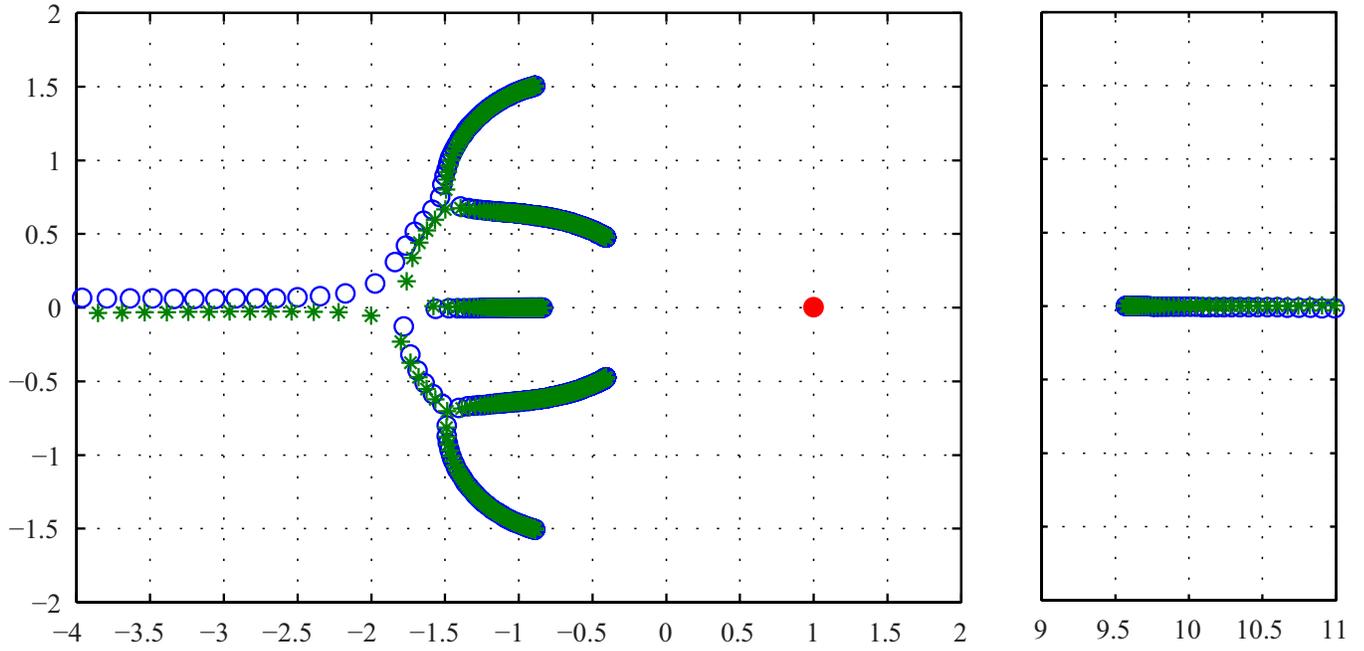}}
           \\
        \subfloat[Transformation of the Stahl's compact set on the feasibility boundary ($P_1=2.6785$)\label{ZIP_bf}]{
        \includegraphics [scale=1,trim=0.0cm 0.cm 0.0cm 0.0cm,clip]{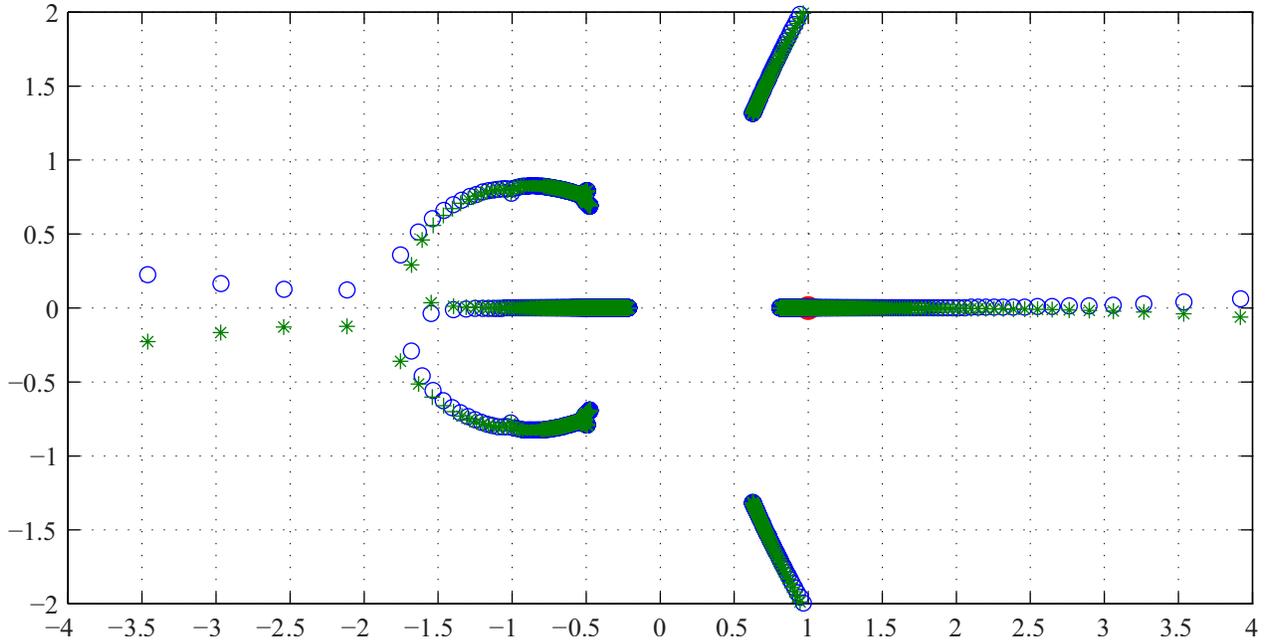}}
         \caption{Zero-pole distribution of PA[1000/1000] depicting the analytic structure of $V_1(z)$  (corresponding to Fig.~\ref{grid2})}.
\end{figure*}

\begin{align}\label{pg1}
&\displaystyle z\sum_{k\in \mathcal{L}[i]} I_{ik}(V_i,S_{ik},m_k,n_k)=\hspace{-2mm}\sum_{k\in \mathcal{N}[i]} V_k(z) Y_{ik} \hspace{4mm} \forall i\in\mathcal{E}
\end{align}

\noindent
where $\mathcal{L}[i]$ is the set of load components of node $i$ each characterized by its $S_0$ and $(m,n)$ and $\mathcal{E}$ is the set of nodes with exponential load components.

Suppose a given node has a constant-current load where $m=n=1$ in~\eqref{EL3} then the corresponding power series relation is given as,

\begin{align}\label{ccl}
&\displaystyle  (\sum_{k=0}^{\infty} x_kz^k)^2 =(S_0^*)^2(\sum_{k=0}^{\infty} c_kz^k)(\sum_{k=0}^{\infty} {d^*_k}z^k)
\end{align}

The power series $I(z)=\sum_{k=0}^{\infty} x_kz^k$ is developed by successive application of the concept of convolution. The coefficient $x_{k}$ is obtained based on $c_j$, $d_j$ and $x_j$ for $j=1,...,k-1$, all given from previous stages, as follows,

\begin{align}\label{ccl2}
&\displaystyle x_{k}=\displaystyle \frac{(S_0^*)^2 \displaystyle \sum_{j=0}^{k}c_{k-j}d^*_j - \displaystyle \sum_{j=1}^{k-1}x_{k-j}x_j}{2x_0}
\end{align}

Note that the order of power series convolution in the right side of~\eqref{ccl2} increases as $m$ and $n$ assume larger values in~\eqref{EL3}. Thus the corresponding formula for obtaining $x_k$ can be much more complicated. The increasing complexity is related to the partitioning of integers. For example to compute $x_{100}$, the 101st coefficient of $I$, for $n=7$, we need to have obtained, beforehand, all partitions of integer 100 with number of parts less than or equal to $2n=14$. One such partition has only a single part which is 100 itself and its corresponding term is $2n x_0^{2n-1}x_{100}$. All other partitions have corresponding terms that are solely expressed in terms of $x_0,x_1,...,x_{99}$, all obtained in previous stages. This task is, however, independent of the problem itself and is purely in the realm of number theory and combinatorics. In fact there are generating algorithms for these partitions~\cite{Andrews}. All that is needed is to somehow compute and store these partitions for relevant combinations of $(m,n)$. We expect that $I$ coefficients are obtained quite efficiently via parallel computing when $m$ and $n$ are small integers but we leave these aspects, including the need for higher numerical precision, for future publications. Here we focus on the ZIP load model, especially its constant-current component.

Figure~\ref{ZIP} shows the Stahl's compact set corresponding to the network of Figure~\ref{grid2} where bus 2 has a constant-current load, i.e. $P_2=-0.50|V_2|$ and $Q_2=-0.30|V_2|$, and $P_1=1.00$. The Pad\'{e} approximated solution of $V_2(z)$ and $I_2(z)$ are $0.7795+j0.3032$ and $-0.5747+j0.0983$ and hence $|I_2|=|S_2|$ as expected. For this operating point the voltage dependency of the load enhances the stability margin as seen by the location of the branch point on the positive real axis at $z_{\text{b}}=9.5700$). This is because as the system is stressed by increased loading and as the voltage at bus 2 declines, the realized load which is proportional to voltage magnitude decreases further and further from the load at nominal voltage. This drop in the actual load is clearly beneficial for the voltage stability of the system. The situation is the opposite when the system is stressed by increasing active power generation. Under this condition where the voltage magnitude of the load is suppressed by increasing the injection of active power a voltage-dependent load would further stress the system by accentuating the effect of increased active power injection. This becomes evident by contrasting~Figure~\ref{ZIP_bf} where bus 2 has a constant-current load and Figure~\ref{Pade_bf} where bus 2 has a constant-power load. Notice that under constant-power load the network can absorb $P_1=2.6785$ whereas this level of generation is no longer feasible under constant-current load as clearly shown by the analytic arc on real axis covering $z=1$. This contrast highlights the significance of appropriate load models for voltage stability studies.


\appendix[On the Stahl's Theory and the rate of Convergence of Pad\'{e} Approximants]

Assume we are given a germ at point $z_0$ of a multi-valued analytic
function~$f$. Typically this means that we are given the following power series,
\begin{equation}
\label{1s}
S(z)\simeq\sum\limits_{k=0}^\infty c_k(z-z_0)^k
\end{equation}
This is the Taylor series expansion of the analytic function developed at $z=z_0$ and its radius of convergence might be
calculated via Cauchy--Hadamard formulae as,
\begin{equation}
\label{2s}
\frac1R=\varlimsup_{k\to\infty}|c_k|^{1/k}.
\end{equation}
Let $S_n(z)\hspace{-1mm}=\hspace{-1mm}\sum_{k=0}^n c_k(z-z_0)^k$ be a partial sum of the power series and $S_\infty(z)$ the limit of $S_n(z)$ as $n\to\infty$.
Then $f$ can be evaluated in the disk $D_R:=\{z\in\CC:|z-z_0|<R\}$
as,
\begin{equation}
\label{3s}
f(z)=S_\infty(z)\quad\text{for}\quad |z-z_0|<R
\end{equation}

Now suppose that $f$ is a multi-valued analytic function on the
Riemann sphere $\myo\CC$ punctured at a finite set $\Sigma=\Sigma_f$ of
the remarkable points of $f$ at least one of which is a branch point of~$f$.
Let $\Sigma=\{b_1,\dots,b_p\}$ and suppose $z_0\notin\Sigma$. The case
when multi-valued function~$f$ is an algebraic function is of great
interest to us. Function $f$ is an algebraic of degree $m$
if there exists an irreducible complex polynomial $P(z,w)$ in two
complex variables $z$ and $w$ and of degree~$m$ in $w$ such that
$P(z,f(z))\equiv0$ for $z\notin\Sigma$. One can invoke Cauchy's argument principle to show that an algebraic function (assuming $m \ge2$) is also a {\em multi-valued} analytic function. Recall that the essential idea of holomorphic embedding is to analytically
continue the germ of voltage phasors from $z=z_0=0$ where the power flow has a trivial
stable solution to $z=1$. So hereafter we assume $z_0=0$. By definition, analyticity of $f$ in the domain $G:=\myo\CC\setminus\Sigma$ implies that a germ of~$f$ given at the point $z=0$ might be continued analytically from $z=0$ to each point $z=a$, $a\notin\Sigma$, along every path
$\gamma$ such that $\gamma$ avoids points of the set $\Sigma$, i.e.
$\gamma\subset\myo\CC\setminus\Sigma$. Since function $f$ is
multi-valued in $G$, two different paths, say $\gamma_1$ and
$\gamma_2$, such that $\gamma_1,\gamma_2\subset G$ and both of
$\gamma_1$ and $\gamma_2$ connect the original point $z=0$ with the end
point $z=a$, might lead to two different germs $f_1$ and $f_2$ of
$f$ at just the same point $z=a$. In others words, one can obtain
$f_1(z)\neq f_2(z)$ in some neighborhood of $a$. This is always the case
when two paths $\gamma_1$ and $\gamma_2$ form a closed curve
$\gamma=\gamma_1\cup\gamma_2$ such that it encircles exactly one branch
point $b_j\in\Sigma$ of $f$. This is an example of a {\em nontrivial
monodromy} of the closed path $\gamma$. Now suppose that we fix
some simply connected subdomain $D$ of $G$ such that $\{0,a\}\subset D$. Recall
that simple connectivity means that $\myo\CC\setminus D$, the complement
of $D$, is a connected set. From this it follows immediately that
for each path $\gamma$ in $D$ its monodromy is trivial. Then from
the classical monodromy theorem it follows that for every two paths
$\gamma_1$ and $\gamma_2$ which both lead from $z=0$ to $z=a$ the corresponding germs $f_1$ and $f_2$ are identical, i.e.
$f_1(z)\equiv f_2(z)$ in some neighborhood of the point $z=a$.

From now on we shall restrict our attention to the case of algebraic functions
only.

Given a germ at $z=0$ of a multi-valued analytic function~$f$  with a
finite set $\Sigma=\{b_1,\dots,b_p\}$ of remarkable points, one can
evaluate the function $f(z)$ for $|z|<R$, where
$R=\min\{|b_j|,j=1,\dots,p\}$, via the equality $f(z)=S_\infty(z)$.
Since the disk $D_R=\{|z|<R\}$ is a simply connected domain, from the
monodromy theorem it follows that for each closed path $\gamma$ from the
disk $D_R$ its monodromy is trivial.

There exists a powerful method to evaluate the analytic function via
its germ, given at the point $z=0$, beyond the boundary of the disk of convergence
of the corresponding power series. In this method which is also classical and is
based on the notion of continued fractions (to be more precise continued $C$-fractions), we start from the given power series and use the
classical Viskovatov algorithm to obtain (under some additional
assumptions of nondegeneracy of a given germ as it is generically always
the case) the formal expansion,\vspace{-2mm}

\begin{equation}
\label{4s}
S(z)\simeq c_0+c_1z+c_2z^2+\dotsb\simeq
c_0+\cfrac{c_1^{(0)}z}{1+\cfrac{c_1^{(1)}z}{1+\cfrac{c_1^{(2)}z}{1+\dots}}}
\simeq C(z)
\end{equation}

This classical approach of evaluating an analytic function from its germ was known, in some partial forms, since the time of Jacobi and Gauss as continued fraction expansion (or J-fraction
expansion named after Jacobi). But at that time it was used to expand only special functions, in particular the hypergeometric functions. They had recognized that J-fraction expansions give the single valued continuation of a germ of a multivalued analytic function from the
origin into an \emph{unknown} domain which is much larger than the initial disk of convergence.

Let us now consider a germ of an algebraic function $f$, developed at $z=0$, and let $C_n(z)$
be the $n$-th truncate of the $C$-fraction~\eqref{4s}, i.e.\vspace{-2mm}
\begin{equation}
\label{5s}
C_n(z)=
c_0+
\cfrac{c_1^{(0)}z}{1+\cfrac{c_1^{(1)}z}{1+\cfrac{c_1^{(2)}z}{\ddots\cfrac{c_1^{(n-2)}z}
{1+c_1^{(n-1)}z}}
}}
\end{equation}
Let $C_\infty(z)$ be the limit of the truncated $C_n(z)$ as $n\to\infty$.
Then a number of fundamental problems arise around the equality (cf.~\eqref{3s})
\begin{equation}
\label{6s}
f(z)=C_\infty(z)
\end{equation}
The main problems are as follows. In what domain $D$ of the complex
variable $z$ and in what sense the equality~\eqref{6s} holds true?
Since all $C_n(z)$ functions are rational in $z$ and thus single-valued on $\myo\CC$, the limit function $C_\infty(z)$ is also single-valued. This is in contrast to that fact that the initial function $f$ is multi-valued. It is well-known
that in general for each $n$ there is a finite number of the so-called
spurious zero-pole pairs of $C_n(z)$ that do not correspond to any singularity
of the given function $f$ (and neither correspond to a pole or a zero of
$f$; see~\cite{Fro69} and also~\cite[Chapter 2, \S\,2.2]{BaGr96}
and~\cite{KoIkSu15b}). Such pairs are usually referred to as `Froissart
doublets'. As $n$ tends
to infinity, pole and zero in such a pair come close to each other. On one hand, they cancel each other as
$n\to\infty$, but on the other, for a fixed $n$, they are distinct
from each other and as $n\to\infty$ they are dense everywhere on the Riemann
sphere $\myo\CC$. For a Riemann surface of genus~$1$ they make some kind
of 'winding of the torus'. By that reason there can not be a pointwise
convergence of $C_n(z)$ to $f(z)$ for $z\in D$, i.e. the pointwise equality
$f(z)=C_\infty(z)$ in $D$ should not be expected at all.

Recall once again that for
an algebraic function $f$ the number of Froissart doublets is finite,
i.e. is independent of $n$, and depends on $f$ only.
For example, let $f$ be given by the algebraic equation $(1-z^2)w^2-1\equiv0$,
i.e. $f(z)=1/\sqrt{1-z^2}$, and let us fix the germ at $z=0$ by the
equality $f(0)=1$. Then all zeros and poles of $C_n(z)$ belong to the
complement $\myo\RR\setminus[-1,1]$ of the closed segment $[-1,1]$ and
its inverses has the limit distribution on the segment $[-1,1]$ that
coincide with Chebyshev measure given by
$\dfrac1\pi\dfrac{dx}{\sqrt{1-x^2}}$, $x\in[-1,1]$. There are no
Froissart doublets in that case and thus the equality
$f(z)=C_\infty(z)$ holds true pointwise for $z\in D:=\myo\CC\setminus
[-1,1]$. In contrast, the equality $f(z)=S_\infty(z)$ holds
true only for the unit disk, i.e. for $|z|<1$.

The problem of equality in~\eqref{6s} $f(z)=C_\infty(z)$ for an arbitrary
multi-valued function with a finite set of branch points
$\Sigma=\{b_1,\dots,b_p\}$ was completely solved by H.~Stahl in
1985--1986 (see~\cite{Sta85a}--\cite{Sta86b},~\cite{Sta97b} and
also~\cite{ApBuMaSu11},~\cite{KoIkSu15b}).

Given a
germ\footnote{That is the convergent power series at the point $z=z_0$.} $f$ of a multi-valued analytic function $f$ with a finite number of
branch points, the seminal Stahl's theorem\footnote{By the reason of
its very general character and the various subjects of complex analysis
involved into the proof of the theorem, it is sometimes considered as
`Stahl's Theory'.} gives the complete answer to the problem of limit
zero-pole distribution of the classical Pad\'{e} approximants and the equality
of $C_\infty$ to~$f$. The keystone of the Stahl's
theorem is the existence of a unique so-called `maximal domain' of holomorphy of a given multi-valued function $f$, i.e. a domain
$D=D(f)\ni 0$ such that the given germ $f$ can be continued as
holomorphic (i.e. analytic and single-valued) function from a
neighborhood of $z=0$ to $D$ (i.e. the function~$f$ is
continued analytically along each path that belongs to $D$). `Maximal'
means that $\partial{D}$ is of `minimal capacity' (with respect to the point $z=0$) among all compact
sets $\partial{G}$ such that $G$ is a domain, $G\ni 0$ and
$f\in \mathscr{H}(G)$. Such `maximal' domain $D$ is unique up to a compact
set of zero capacity. Compact set $S=S(f):=\partial{D}$ is now called the
`Stahl's compact set' or the `Stahl's $S$-compact set' and $D$ is called
the `Stahl's domain'. The crucial properties of $S$ for the Stahl's theorem
to be true are the following: the complement $D=\myo\CC\setminus{S}$ is
a domain, $S$ consists of a finite number of analytic arcs, and finally $S$ possesses some special property of
`symmetry'\footnote{Compact sets of such type are usually referred to
as `$S$-compact sets' or `$S$-curves',
see~\cite{Rak12},~\cite{KuSi15}.}.

From the Stahl's theorem it follows that the limit points of the zero and
pole distributions of rational functions $C_n(z)$ as $n\to\infty$ exists
and coincides with a unique so-called probability equilibrium measure for
the Stahl's compact set $S$. From the numerical point of view this means,
first, that zeros and poles of $C_n(z)$ are attracted as $n\to\infty$ to
the Stahl's compact set $S$. Second, they accumulate to each\footnote{
To be more precise, to each `active' branch point.  All the so-called `inactive'
branch points are hidden on the other `nonphysical' sheets of the Riemann
surface of the given function; see~\cite{Sta12}.}
branch point
$b_k\in S\cap\Sigma$ of $f$ with a density similar to that of the Chebyshev
measure $ \frac1\pi\frac{dx}{\sqrt{1-x^2}}$ for $S=[-1,1]$ at the end
points $\pm1$.

On the Stahl's compact set $S$ there are also a finite number of the so-called
Chebotarev's points that do not correspond to any branch point of
the initial function $f$ and at these points the equilibrium measure of $S$
has a zero density similar to that of the measure $\sqrt{1-x^2}\,dx$ at the end points $\pm1$. These Chebotarev's points are the
transcendental parameters of the problem, i.e. they can not be recovered from the
branch points of a given function over elementary functions but only over
transcendental functions. For example in the case of the function $f(z)=(b_1-z)^{1/3}(b_2-z)^{1/3}(b_3-z)^{-2/3}$ with the three branch points $b_1,b_2,b_3$ this means that the Chebotarev's point $v$ is uniquely determined from the condition that the both periods of the Abelian integral are purely imaginary:
\begin{equation}
\int^z\sqrt{\frac{v-\eta}{B_3(\eta)}}\,\frac{d\eta}{\eta}
\label{a3}
\end{equation}
\noindent
where $B_3(\eta):=\prod_{j=1}^3(b_j-\eta)$. Chebotarev's points, jointly with the branch points
of $f$,  determine the Stahl's two-sheeted hyperelliptic Riemann
surface associated with $f$. Thus numerically in a neighborhood of a Chebotarev's point, zeros and poles of
$C_n(z)$ are very sparse compared to the number $n$. It
might be concluded that zeros and poles of $C_n(z)$ as $n$ becomes large
enough, eventually will recover numerically the complete structure of the Stahl's compact
set $S$ and the Stahl's domain $D$.

Finally in the Stahl's theorem it is proved that the equality
$f(z)=C_\infty(z)$ holds true in the Stahl's domain $D=D(f)$ not
pointwise but `in capacity'. The convergence in capacity inside the Stahl's
domain $D$ (recall that $f\in\HH(D)$) means that for every compact set
 $K\subset D$ and for every small $\varepsilon>0$ the following holds,

\vspace{-1mm}
\begin{equation}
\label{cap}
\mcap\{z\in K: |f(z)-[n/n]_f(z)|\geq\varepsilon>0\}\to0, n\to\infty, z\in D
\end{equation}

\noindent where $\text{cap}(\cdot)$, is the logarithmic
capacity~\cite{Saff97}. The only reason for this specific mode of
convergence is the existence of a finite number of Froissart doublets. In fact
the truncated $C_n(z)$ of the $C$-fraction in~\eqref{3s} gives a very good
numerical approximation of $f(z)$ in all the points $z$ of the Stahl's
domain up to a finite number of `wandering' Froissart doublets.

The rate of the convergence in~\eqref{cap} is completely characterized by the equality
\begin{equation}
|(f-[n/n]_f)(z)|^{1/n}\overset{\mcap}\longrightarrow
e^{-2g^{}_S(z,0)}, n\to\infty, z\in D
\label{green}
\end{equation}
\noindent
where $g_S(z,0)$ is the Green's function of the domain $D$
with a logarithmic singularity at the point $z=0$. $g_S(z,0)$ is defined in relation to a given branch point $b_j$ in the following way,
\begin{equation}
g_S(z,0):=\Re
\int_{b_j}^z\sqrt{\frac{v-\eta}{B_3(\eta)}}\,\frac{d\eta}{\eta}
\end{equation}

Thus the rate of convergence at a given point $z\in D$ depends on the value of the Green's function $g_S(z,0)$ for the domain $D \ni 0$ at that point. The closer the point $z$ gets to the boundary $S:=\partial{D}$ of $D$, the smaller the rate of convergence becomes. In the disk $D_R$ the convergence of $C_n(z)$ to $f(z)$ is much faster than the
convergence of partial sums $S_n(z)$. Finally we should mention that in generic cases for an
algebraic function $f$ given by polynomial equation $P(z,f(z))\equiv0$, the
corresponding Stahl's compact set is stable under small perturbations of
complex coefficients of the polynomial $P$.


%
%
%
\pagebreak


\end{document}